\documentclass{aa}  

\usepackage{graphicx}
\usepackage{txfonts}
\usepackage[hidelinks]{hyperref}

\begin{document}

\title{A strategy for astrometric quasar selection with Gaia DR3 and its application to discover new luminous $z>2$ quasars}\titlerunning{Astrometric quasar selection with Gaia DR3}

\author{J. Li \inst{1,2}\email{djw465@alumni.ku.dk} 
\and J. P. U. Fynbo \inst{1,2}\corrauth{jfynbo@nbi.ku.dk}}

\institute{Cosmic Dawn Center (DAWN), Denmark
\and Niels Bohr Institute, University of Copenhagen, Jagtvej 128, 2200 Copenhagen N, Denmark}

\date{Received September XX, 2026}

\abstract
   {We provide a cookbook for a purely astrometric selection of quasars with Gaia DR3 and provide an example of its use for the identification of 
   very luminous $z>2$ quasars.}
   {We analysed the Gaia DR3 astrometry of confirmed quasars and stationary sources. The proper motion signal-to-noise ratio  $\mu/\sigma_{\mu}$ follows a Rayleigh-like distribution. We also show the importance of 
   properly taking the correlation between proper motions in right ascension and declination into account, as well as the non-linear error propagation. We adopted the Mahalanobis distance to the null proper-motion hypothesis. A scaled $\chi^2$ distribution with $k=2$ and scale $s=1.1485$ was fitted to Gaia-CRF3; the corresponding $\pm2\sigma$ Gaussian-equivalent threshold is $D^2_{M(\mu)}<7.097$. The final criteria combine this with the parallax condition and $\mathrm{ruwe}<1.4$.}
   {When we applied the new criteria to the SDSS DR16Q confirmed quasars, the recovery rate was $90.25\%$.}
   {The suggested purely astrometric selection significantly improves completeness and purity compared to other strategies used in the literature, reaching a quasar purity of approximately $95\%$ ($\sim2\sigma$ Gaussian-equivalent). It can be used to improve the selection purity and completeness for future quasar surveys. As an example, we used the suggested strategy to locate 24 very bright $z>2$ quasars without 
   redshifts in the literature.}
   {Our proposed astrometric selection strategy recovers 95\% of the known quasars. Its potential use for new discoveries
   was demonstrated by the identification of 24 very luminous $z>2$ quasars. Of these, 10 are new discoveries, 
   and for an additional 7, we corrected the redshifts determined by other methods.}


\maketitle
\nolinenumbers

\section{Introduction}

Since quasars, or quasi-stellar objects~(QSOs), were discovered in the early 1960s \citep{schmidt3C273StarLike1963}, many quasar surveys have been carried out, and about one million quasars are known today. To obtain a complete understanding of quasars, it is desirable to construct quasar samples that are as complete as possible in multiple ways, in order to mitigate the limitations caused by selection effects.

Many studies have investigated the incompleteness of quasar samples resulting from different selection methods. In the radio band, \citet{kellermannVLAObservationsObjects1989,ivezicOpticalRadioProperties2002} reported that only $\sim 10\%$ of the quasars are radio-loud. Later, \citet{richardsSpectroscopicTargetSelection2002,richardsEfficientPhotometricSelection2004} discussed the colour-selection effects in the optical photometric selection of quasars from the Sloan Digital Sky Survey~(SDSS). It is now well established that these samples miss a substantial number of quasars \citep[e.g.][]{krawczykMININGDUSTTYPE2015}.

In an earlier study, \citet{heintzStudyPurelyAstrometric2015} introduced a new method for selecting quasar candidates purely based on astrometric data that identifies sources on the sky with zero parallax and zero proper motion within the errors as potential quasars \citep[see also][]{1986PASP...98..285K,2002A&A...392..851M}. Subsequent studies, for instance \citet{heintzUnidentifiedQuasarsStationary2018,heintzSpectroscopicClassificationComplete2020,Geier2019,Fynbo2020}, demonstrated that this is a viable approach for recovering quasars, including quasars that are missed by traditional optical selection criteria. 

We employ a proper treatment of the statistics of astrometric measurements and outline a cookbook for using
astrometry from the Gaia mission as a quasar selection technique. To this end, we describe the Gaia data in Sect.~\ref{sec:pre}. In 
Sect.~\ref{sec:sol} we describe our proposed new astrometric selection strategy for quasars. In Sect.~\ref{sec:res}
we validate our proposed strategy by analysing the Sloan Digital Sky Survey Quasar Catalog: Sixteenth Data Release~(SDSS DR16Q) from \citet{lykeSloanDigitalSky2020}. In Sect.~\ref{sec:application}
we apply our proposed strategy to select new luminous $z>2$ quasars in the northern sky, and use spectroscopic data from the Spectro-Photometer for the History of the Universe, Epoch of Reionization, and Ices Explorer (SPHEREx), the Large Sky Area Multi-Object Fiber Spectroscopic Telescope (LAMOST), the Dark Energy Spectroscopic Instrument (DESI), and SDSS to verify their classifications and refine their redshifts. 
Finally, in Sect.~\ref{sec:dis} we offer our discussion and conclusions.

\section{\textit{Gaia} data}
\label{sec:pre}

The \textit{Gaia} mission \citep{gaiacollaborationGaiaMission2016}, launched by the European Space Agency~(ESA) in December 2013, is a powerful astrometric mission. The mission is designed to make a three-dimensional map of our Galaxy and to provide unprecedented measurements of positions, proper motions, and parallaxes for more than one billion stellar systems down to $G \approx 20.7$ mag during its ten-year mission lifetime.

\subsection{\textit{Gaia} DR3}

\textit{Gaia} Data Release 3~(\textit{Gaia} DR3) is the third public \textit{Gaia} data release, published on 13 June 2022 by \citet{gaiacollaborationGaiaDataRelease2023}. These data were produced by the Gaia Data Processing and Analysis Consortium from the first 34 months of observations. The \textit{Gaia} DR3 catalogue contains the same source list, celestial positions, proper motions, parallaxes, and broadband photometry in the $G$, $G_{BP}$, and $G_{RP}$ passbands as the Early Third Data Release (Gaia EDR3) presented by \citet{gaiacollaborationGaiaEarlyData2021a}.

\textit{Gaia} DR3 provides celestial positions and apparent brightness in $G$ for over 1.8 billion sources. For about 1.5 billion of these sources, parallaxes and proper motions are also published. All \textit{Gaia} data used in this work were obtained from the \textit{Gaia} Archive website\footnote{\url{https://gea.esac.esa.int/archive/}} or the VizieR catalogue service\footnote{\url{https://vizier.cds.unistra.fr/viz-bin/VizieR}}.

\subsection{Previous strategies for astrometric selection}

In previous studies such as \citet{heintzUnidentifiedQuasarsStationary2018}, two main criteria were applied for the astrometry selection.
The first criterion was that the measured parallax $\varpi$, after correction for the median parallax bias, be consistent with zero within twice the formal uncertainty $\sigma_{\varpi}$,
\begin{equation}
    \left|\frac{\varpi+0.017\,\mathrm{mas}}{\sigma_{\varpi}}\right|<2
    \label{eq:parallax_over_error}.
\end{equation}
This translates into the Astronomical Data Query Language~(ADQL) query
{\small\begin{verbatim}
abs((parallax+0.017)/parallax_error) <= 2.
\end{verbatim}}
This correction accounts for the median parallax bias of $-0.017$ mas reported by \citet{fabriciusGaiaEarlyData2021}, but neglects the complex dependence of the parallax bias on magnitude and other parameters discussed by \citet{lindegrenGaiaEarlyData2021a}.

The second criterion is that the proper motion $\mu$ should be consistent with zero within twice its uncertainty $\sigma_{\mu}$,
\begin{equation}
     \frac{\mu}{\sigma_{\mu}}=
     \sqrt{\left(\frac{\mu_{\alpha*}}{\sigma_{\mu_{\alpha*}}}\right)^2
     + \left(\frac{\mu_{\delta}}{\sigma_{\mu_{\delta}}}\right)^2} < 2
     \label{eq:old_pm}.
\end{equation}
The corresponding ADQL query is
{\small
\begin{verbatim}
(power(pmra/pmra_error,2) + 
power(pmdec/pmdec_error,2)) <= 4.
\end{verbatim}}

In our initial prediction, this set of criteria should select stationary objects at the $2\sigma$ level on the sky. However, when we cross-checked with the SDSS DR16Q quasar catalogue \citep{lykeSloanDigitalSky2020} and the \textit{Gaia} DR3 QSO candidate catalogue \citep{gaiacollaborationGaiaDataRelease2023a,delchambreGaiaDataRelease2023}, we found that the number of missing targets was nearly three times our expectation.

\subsection{The statistical distribution of the proper motion}

\citet{lindegrenAstrometricCoreSolution2012} described how the \textit{Gaia} team solves the astrometric parameters of objects from the observations. It is straightforward to see that the proper motion is measured as two components, $\mu_{\alpha*}$ and $\mu_{\delta}$. These two components have their own uncertainties, $\sigma_{\mu_{\alpha*}}$ and $\sigma_{\mu_{\delta}}$. As a result, $\mu_{\alpha*} / \sigma_{\mu_{\alpha*}}$ and $\mu_{\delta} / \sigma_{\mu_{\delta}}$ can theoretically be treated as two independent standard normal variables. In our original selection process, we approximated ${\mu}/{\sigma_{\mu}}$ from Eq.~\ref{eq:old_pm} as a normal variable to estimate the number of missing candidates, but this non-linear combination inherently does not follow a normal distribution. ${\mu}/{\sigma_{\mu}}$ follows a Rayleigh distribution. The Rayleigh distribution describes the magnitude of a two-dimensional vector whose orthogonal components follow independent zero-mean normal distributions.

\begin{figure}[ht!]
    \centering
    \includegraphics[width=\hsize]{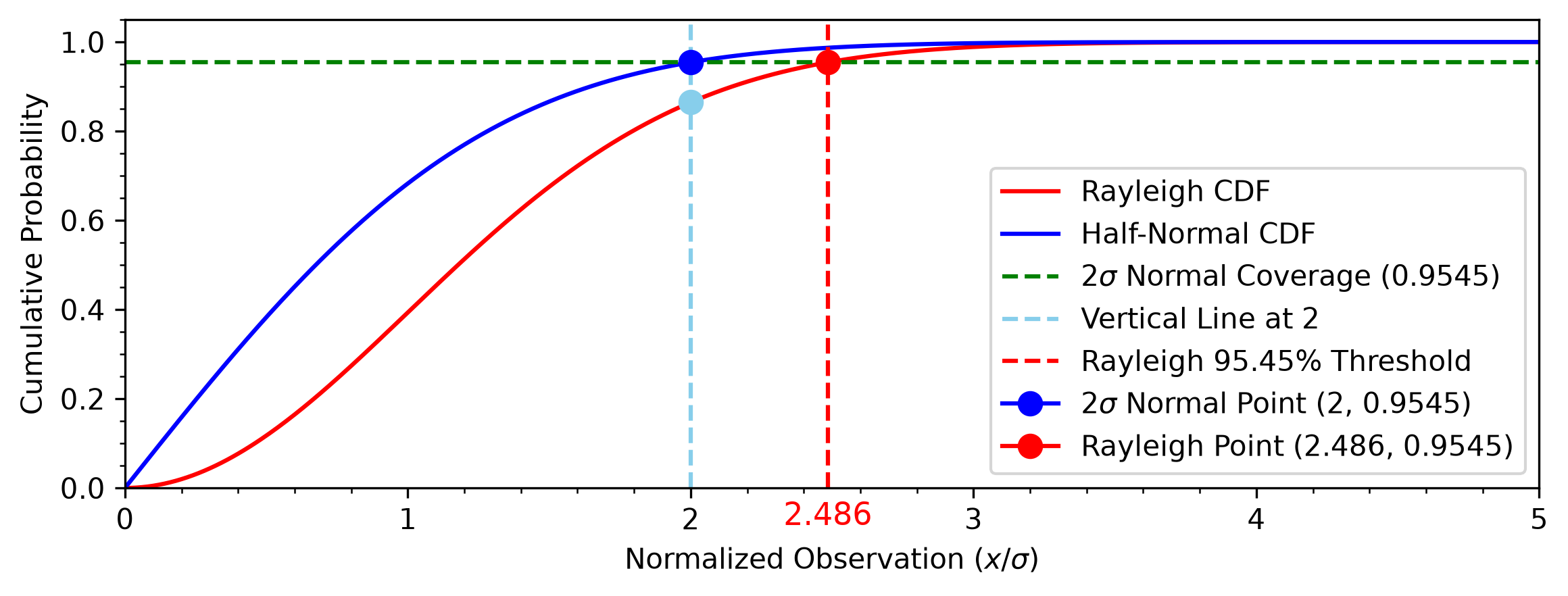}
    \caption{CDFs of the half-normal and Rayleigh distributions in units of the standard deviation. The dashed green line indicates the probability that measurements fall within $\pm 2\sigma$ of a standard normal distribution.}
    \label{fig:ray_norm_cdf}
\end{figure}

For a standard normal distribution, approximately $95.45\%$ of the values fall within the $\pm 2\sigma$ interval, corresponding to an expected miss rate of $4.55\%$. Fig.~\ref{fig:ray_norm_cdf} compares the cumulative distribution functions (CDFs) of these two distributions. When we apply a threshold of 2 to the Rayleigh distribution, it captures only about $86.47\%$ of the measurements (an approximate miss rate of $13.53\%$), resulting in a data loss nearly three times higher than expected. Consequently, a larger threshold is required to retain a similar proportion of the data.

In practice, \citet{fabriciusGaiaEarlyData2021} reported that proper motion uncertainties are underestimated in the catalogue. Therefore, the standard deviation of the actual distribution exceeds 1. By fitting a Rayleigh distribution to the 386\,478 confirmed quasars in the SDSS DR16Q quasar catalogue\footnote{To minimise contamination from sources such as local Seyfert galaxies, only objects with $z>0.5$ are selected. Confirmed quasars from the SDSS DR16Q catalogue (\texttt{QSO} flag = 1 in the VizieR copy) were cross-matched with Gaia DR3 astrometric data through the CDS cross-match service (Strasbourg) using a $1\arcsec$ matching radius. The final catalogue contains 386\,478 objects. \label{fn:sdss-dr16q}} \citep{lykeSloanDigitalSky2020}, we obtained a standard deviation of $\sigma \approx 1.04$, consistent with the value reported by \citet{fabriciusGaiaEarlyData2021}. Under these empirical conditions, a threshold of 2 yields a capture rate of only $83.36\%$. This translates into an approximate $15.64\%$ miss rate, which is nearly four times ($\approx 3.44$ times) higher than the theoretical expectation.

\begin{figure}[ht!]
    \centering
    \includegraphics[width=\hsize]{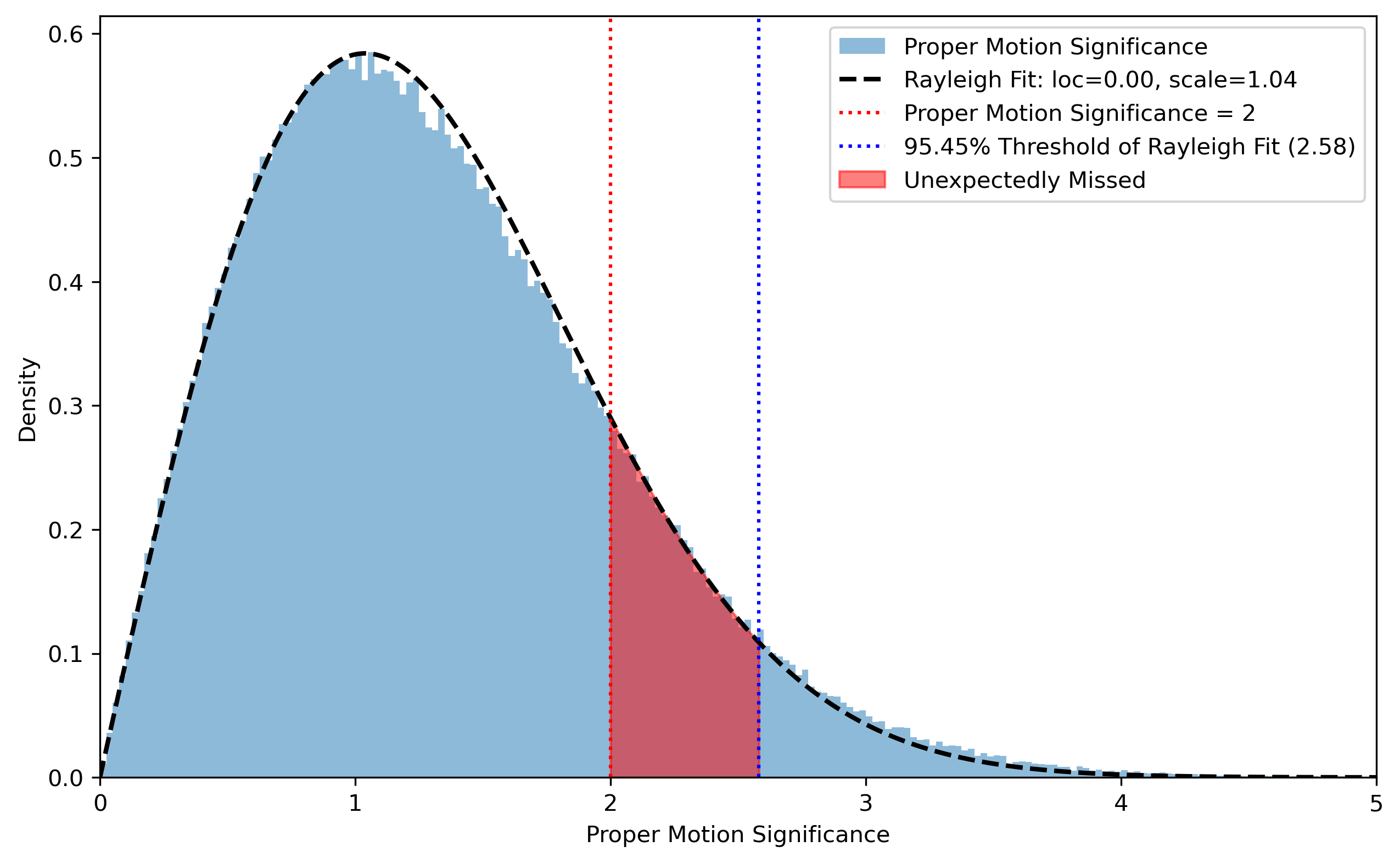}
    \caption{Distribution of the total proper motion criterion in Eq.~\ref{eq:old_pm} for the SDSS DR16Q sample\footref{fn:sdss-dr16q} with 386\,478 confirmed quasars. The solid black line represents the best-fit Rayleigh distribution, and the shaded red area highlights the fraction of candidates that is excluded for a threshold of ${\mu}/{\sigma_{\mu}}<2$.}
    \label{fig:sdss_qso_ray}
\end{figure}

\subsection{Error propagation of the total proper motion in the combination of proper motion in RA and Dec}

Although the method in Eq.~\ref{eq:old_pm} is effective for detecting apparent motion against the celestial background, its application is largely limited to identifying objects with high proper motion or removing background objects. In the following, we examine these limitations and their implications.

First, in practice, measurements of the two proper-motion components $\mu_{\alpha*}$ and $\mu_{\delta}$ are not completely independent. In the Gaia EDR3 table, a column is named \texttt{pmra\_pmdec\_corr}. It gives the correlation $\rho_{\mu_{\alpha*}\mu_{\delta}}$ between the proper motion in right ascension $\mu_{\alpha*}$ and the proper motion in declination $\mu_{\delta}$.

\begin{figure}[ht!]
  \centering
  \includegraphics[width=\hsize]{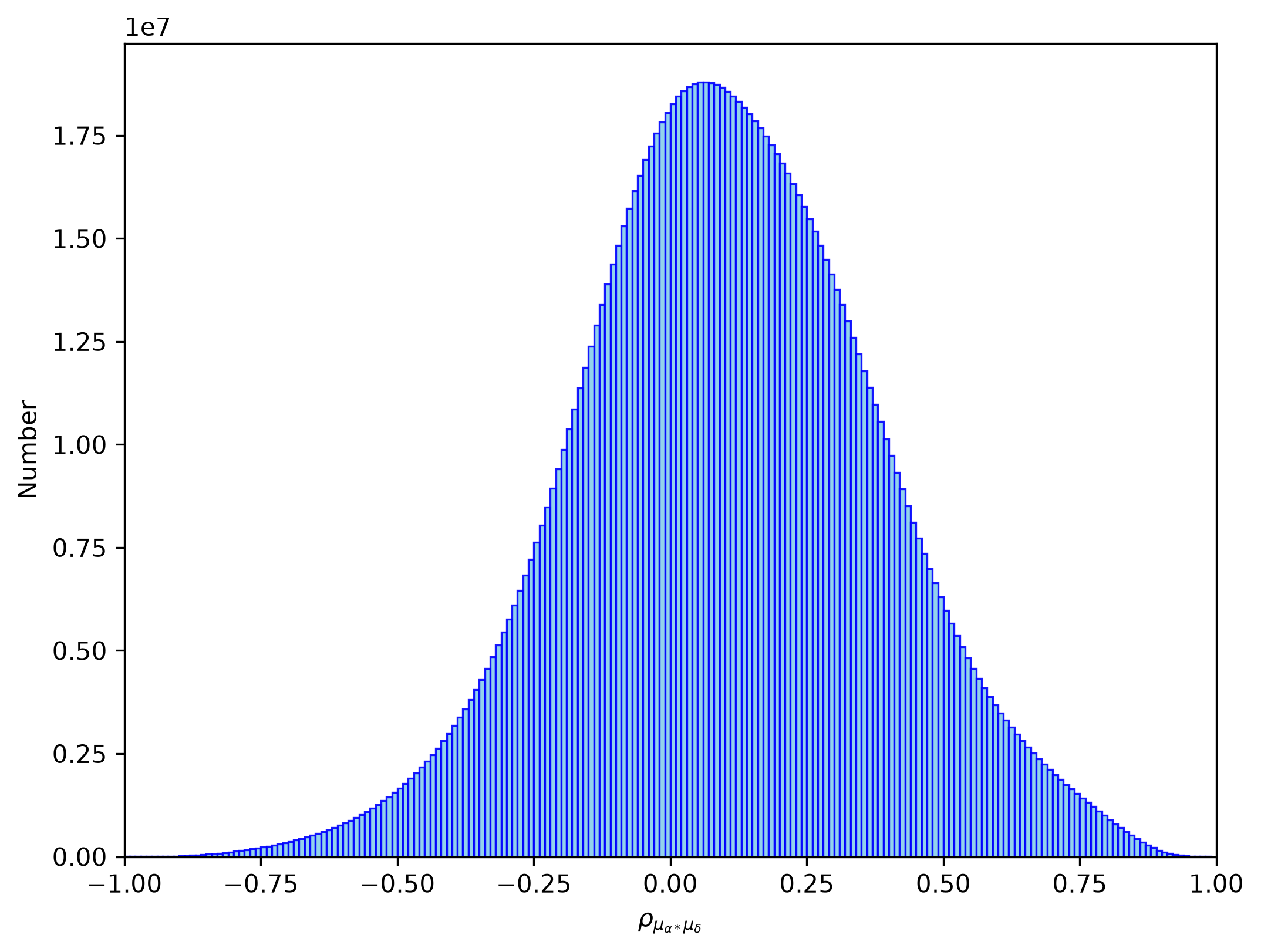}
  \caption{Histogram of $\rho_{\mu_{\alpha*}\mu_{\delta}}$, the correlation between the proper motion in right ascension and the proper motion in declination.}
  \label{fig:pmra_pmdec_corr_dist}
\end{figure}

We examined all $\rho_{\mu_{\alpha*}\mu_{\delta}}$ values for sources with low parallax (consistent with zero at the $2\sigma$ level) over the whole sky in \textit{Gaia} DR3. This sample contained 1\,262\,889\,940 sources. Figure \ref{fig:pmra_pmdec_corr_dist} shows the distribution of these values.

\begin{figure}[ht!]
    \centering
    \includegraphics[width=\hsize]{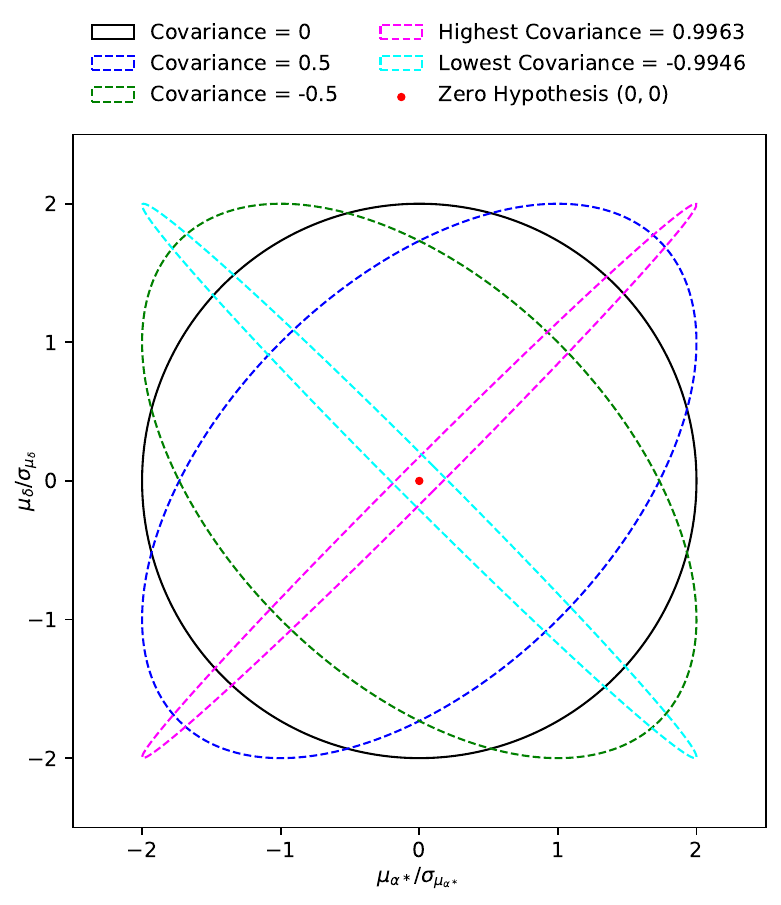}
    \caption{$2\sigma$ error ellipses for the zero hypothesis, showing the effect of the correlation coefficient on the size and shape of the $2\sigma$ area.}
    \label{fig:error_ellipses}
\end{figure}

To comprehensively evaluate how the correlation affects the error propagation for the total proper motion of our targets, we extended this uncertainty analysis to two-dimensional space. 
As illustrated in Fig.~\ref{fig:error_ellipses}, under the ideal assumption that the proper motion measurements in the two directions are statistically independent, our selection can be visualised as identifying targets whose normalised proper motions ($\mu_{\alpha*}/\sigma_{\mu_{\alpha*}}$, $\mu_{\delta}/\sigma_{\mu_{\delta}}$) fall within the circular $2\sigma$ boundary (black circle). However, in practical astrometry, the correlation between the two components is non-negligible. This covariance transforms the selection area into an error ellipse whose shape and orientation depend directly on the correlation coefficient. For illustration, we plot the $2\sigma$ boundary for correlations of $\pm0.5$, $0.9963$ (the highest value in the sample) and $-0.9946$ (the lowest value in the same sample).

Beyond neglecting the correlation, Eq.~\ref{eq:old_pm} has another problem: the uncertainty $\sigma_{\mu}$ is not accurate. To explain this, we begin with the definition of the total proper motion. The total proper motion can be written as
\begin{equation}
    \mu=\sqrt{{\mu_{\alpha*}}^2+{\mu_{\delta}}^2}.
    \label{eq:pm_tot}
\end{equation}

This is a non-linear combination of two components. To obtain the uncertainty $\sigma_{\mu}$, we used a Taylor expansion, also known as the delta method. The first-order approximation of the variance $\sigma_{\mu}^2$ is
\begin{equation}
\begin{split}
    \sigma_{\mu}^2 &\approx \left( \frac{\partial \mu}{\partial \mu_{\alpha*}} \right)^2 \sigma_{\mu_{\alpha*}}^2 + \left( \frac{\partial \mu}{\partial \mu_{\delta}} \right)^2 \sigma_{\mu_{\delta}}^2 +2 \left( \frac{\partial \mu}{\partial \mu_{\alpha*}} \right) \left( \frac{\partial \mu}{\partial \mu_{\delta}} \right) \text{cov}(\mu_{\alpha*}, \mu_{\delta})
    \\
    &\approx \frac{1}{\mu^2}({{\mu_{\alpha*}}^2\sigma^2_{\mu_{\alpha*}}}+{{{\mu_{\delta}}^2\sigma^2_{\mu_{\delta}}}+2\rho_{\mu_{\alpha*}\mu_{\delta}}\mu_{\alpha*}\mu_{\delta}\sigma_{\mu_{\alpha*}}\sigma_{\mu_{\delta}}} ) 
\end{split}
\label{eq:pm_sig_corr}.
\end{equation}

Thus, the corrected version of the criterion $\mu/\sigma_{\mu}$ is
\begin{equation}
\frac{\mu}{\sigma_{\mu}}\approx\frac{{\mu}^2}{\sqrt{{{\mu_{\alpha*}}^2\sigma^2_{\mu_{\alpha*}}}+{{{\mu_{\delta}}^2\sigma^2_{\mu_{\delta}}}+2\rho_{\mu_{\alpha*}\mu_{\delta}}\mu_{\alpha*}\mu_{\delta}\sigma_{\mu_{\alpha*}}\sigma_{\mu_{\delta}}}}}.
\label{eq:sn_pm_new}
\end{equation}

However, owing to the limitations of the Taylor expansion, this approximation is strongly biased when \(\mu_{\alpha*}\) and \(\mu_{\delta}\) are near zero. Unfortunately, our goal is to select stationary objects, for which \(\mu_{\alpha*}\) and \(\mu_{\delta}\) are theoretically zero. Therefore, we cannot ignore the \(O(\sigma^2)\) bias of the delta method. Thus, as discussed above, a more accurate statistical method is needed to quantify the deviation of the proper motion from zero.

\section{Proposed strategy}
\label{sec:sol}

To address these problems, we reviewed several studies based on the \textit{Gaia} DR3 catalogue \citep{fabriciusGaiaEarlyData2021,gaiacollaborationGaiaEarlyData2022}. From these studies, we identified two useful tools that helped us to avoid the aforementioned issues and enabled a more reliable target selection: the Mahalanobis distance and renormalised unit weight error~(\texttt{ruwe}).

\subsection{Mahalanobis distance}
Fundamentally, our selection criterion is based on measuring the statistical distance of the proper motion from the null hypothesis (\(\boldsymbol{\mu} = \boldsymbol{0}\)). Our proposed method builds on this framework.

The previous criterion in Eq.~\ref{eq:old_pm} is mathematically equivalent to calculating the Euclidean distance in the normalised parameter space,
\begin{equation}
    D_{E(\boldsymbol{x},\boldsymbol{y})} = \sqrt{(\boldsymbol{x}-\boldsymbol{y})^2} \implies D_{E(\mu)} = \sqrt{\left(\frac{\mu_{\alpha*}}{\sigma_{\mu_{\alpha*}}}\right)^2 + \left(\frac{\mu_{\delta}}{\sigma_{\mu_{\delta}}}\right)^2}.
    \label{eq:e_dist}
\end{equation}

However, this Euclidean approach assumes that the measurements are fully independent and neglects the cross-correlation between the two proper-motion components. To rigorously account for the covariance, the Mahalanobis distance is the optimal metric. It is generally defined as
\begin{equation}
    D_{M(\boldsymbol{x},\boldsymbol{y})} = \sqrt{(\boldsymbol{x}-\boldsymbol{y})^T \Sigma^{-1}(\boldsymbol{x}-\boldsymbol{y})},
    \label{eq:m_dist}
\end{equation}
where $\Sigma$ represents the covariance matrix. 

For the selection, we set $\boldsymbol{x}=(\mu_{\alpha*}, \mu_{\delta})$ and the null hypothesis $\boldsymbol{y}=(0,0)$. The distance is
\begin{equation}
\begin{split}
    D_{M(\mu)}&=\sqrt{(\mu_{\alpha*}, \mu_{\delta})^T\Sigma^{-1}(\mu_{\alpha*}, \mu_{\delta})}, \\ 
    \Sigma&=
    \begin{pmatrix}
    \sigma^2_{\mu_{\alpha*}} & \rho_{\mu_{\alpha*}\mu_{\delta}} \sigma_{\mu_{\alpha*}} \sigma_{\mu_{\delta}}  \\
    \rho_{\mu_{\alpha*}\mu_{\delta}} \sigma_{\mu_{\alpha*}} \sigma_{\mu_{\delta}} & \sigma^2_{\mu_{\delta}}
    \end{pmatrix}
\end{split}
    \label{eq:m_dist_pm}.
\end{equation}

By substituting the covariance matrix into Eq.~\ref{eq:m_dist_pm}, we obtained the squared Mahalanobis distance,
\begin{equation}
    D^2_{M(\mu)} = \frac{1}{1-\rho^2_{\mu_{\alpha*}\mu_{\delta}}} \left[ \left(\frac{\mu_{\alpha*}}{\sigma_{\mu_{\alpha*}}}\right)^2 + \left(\frac{\mu_{\delta}}{\sigma_{\mu_{\delta}}}\right)^2 - 2\rho_{\mu_{\alpha*}\mu_{\delta}} \left(\frac{\mu_{\alpha*}}{\sigma_{\mu_{\alpha*}}}\right) \left(\frac{\mu_{\delta}}{\sigma_{\mu_{\delta}}}\right) \right].
    \label{eq:m_dist_sq}
\end{equation}
Clearly, the squared Mahalanobis distance $D^2_{M(\mu)}$ is dimensionless.

Theoretically, the squared Mahalanobis distance follows a $\chi^2$ distribution with $k$ degrees of freedom. For $D^2_{M(\mu)}$, $k=2$. 

\begin{figure}[ht!]
    \centering
    \includegraphics[width=\hsize]{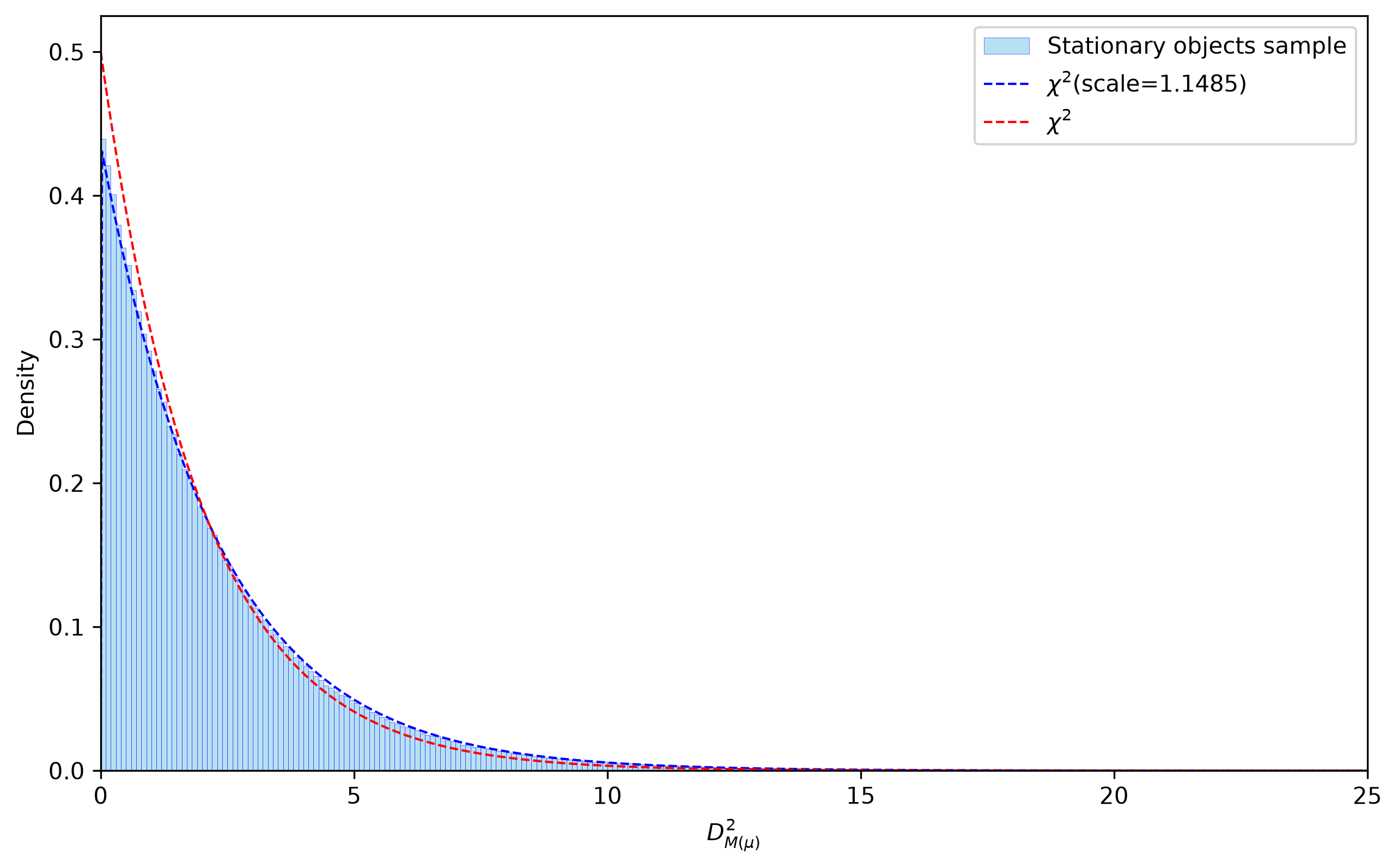}
    \caption{Distribution of $D^2_{M(\mu)}$ for the sample. The dashed red line shows the probability density function~(PDF) of the standard two-degree-of-freedom $\chi^2$ distribution, and the dashed blue line shows the PDF of the fitted distribution with $s=1.1485$.}
    \label{fig:m_fit}
\end{figure}

However, because the uncertainties are underestimated, the standard two-degree-of-freedom \(\chi^2\) distribution in real data does not perfectly fit the \(D^2_{M(\mu)}\) distribution of the stationary-object sample. To address this issue, we used a scaled instead of the standard \(\chi^2\) distribution to describe the empirical distribution. In the SciPy parametrisation, this model can be written as
\[
D^2_{M(\mu)} = s\cdot X,\qquad X \sim \chi^2_{k=2},
\]
where \(s\) is the scale parameter, or equivalently,
\[
f(D^2_{M(\mu)}) = \frac{1}{s} f_{\chi^2_{k=2}}\!\left(\frac{D^2_{M(\mu)}}{s}\right).
\]
To obtain the optimal scale value, we fitted the empirical sample using the maximum likelihood estimation (MLE). The sample we used was the third realisation of the Gaia Celestial Reference Frame \citep[\textit{Gaia}-CRF3;][]{gaiacollaborationGaiaEarlyData2022}. It contained \(1\,614\,173\) objects after those with \(D^2_{M(\mu)}>25\) were removed to mitigate long-tail effects. As a result, we found that a scale of $s=1.1485$ describes the empirical sample best, as shown in Fig.~\ref{fig:m_fit}.

With this distribution, we can define thresholds for \(D^2_{M(\mu)}\) corresponding to specific quantile levels:

\begin{table}[ht!]
\centering
\caption{Thresholds of \(D^2_{M(\mu)}\) for selected quantile levels.}
\label{table:dm}
\begin{tabular}{cc}
\hline\hline
Quantile level & \(D^2_{M(\mu)}\) threshold \\
\hline
\(95\%\)     & 6.882  \\
\(\pm 2\sigma\) & 7.097  \\
\(\pm 3\sigma\) & 13.586 \\
\hline
\end{tabular}
\par\vspace{2pt}
\footnotesize  \(\pm 2\sigma\) and \(\pm 3\sigma\) denote Gaussian-equivalent thresholds.
\end{table}

To validate the effectiveness of this method, we constructed three reference samples comprising galaxies, stars, and quasars for comparison. The galaxy and stellar samples were drawn from the Seventeenth Data Release of SDSS \citep[SDSS DR17;][]{abdurroufSeventeenthDataRelease2022}, using the automatic spectroscopic classifications provided therein. The quasar sample was derived from the SDSS DR16Q catalogue \citep{lykeSloanDigitalSky2020}, which contains objects that were visually inspected and classified as QSOs, broad absorption line~(BAL) QSOs, and blazars. After cross-matching these datasets with Gaia DR3 using a $1\arcsec$ search radius and removing sources with only $3p$ solutions, our final reference samples consisted of
\begin{itemize}
    \item galaxies in SDSS DR17: 54\,035 objects,
    \item stars in SDSS DR17: 615\,304 objects,
    \item quasars in SDSS DR16Q: 168\,060 objects.
\end{itemize}

\begin{figure}[ht!]
    \centering
    \includegraphics[width=\hsize]{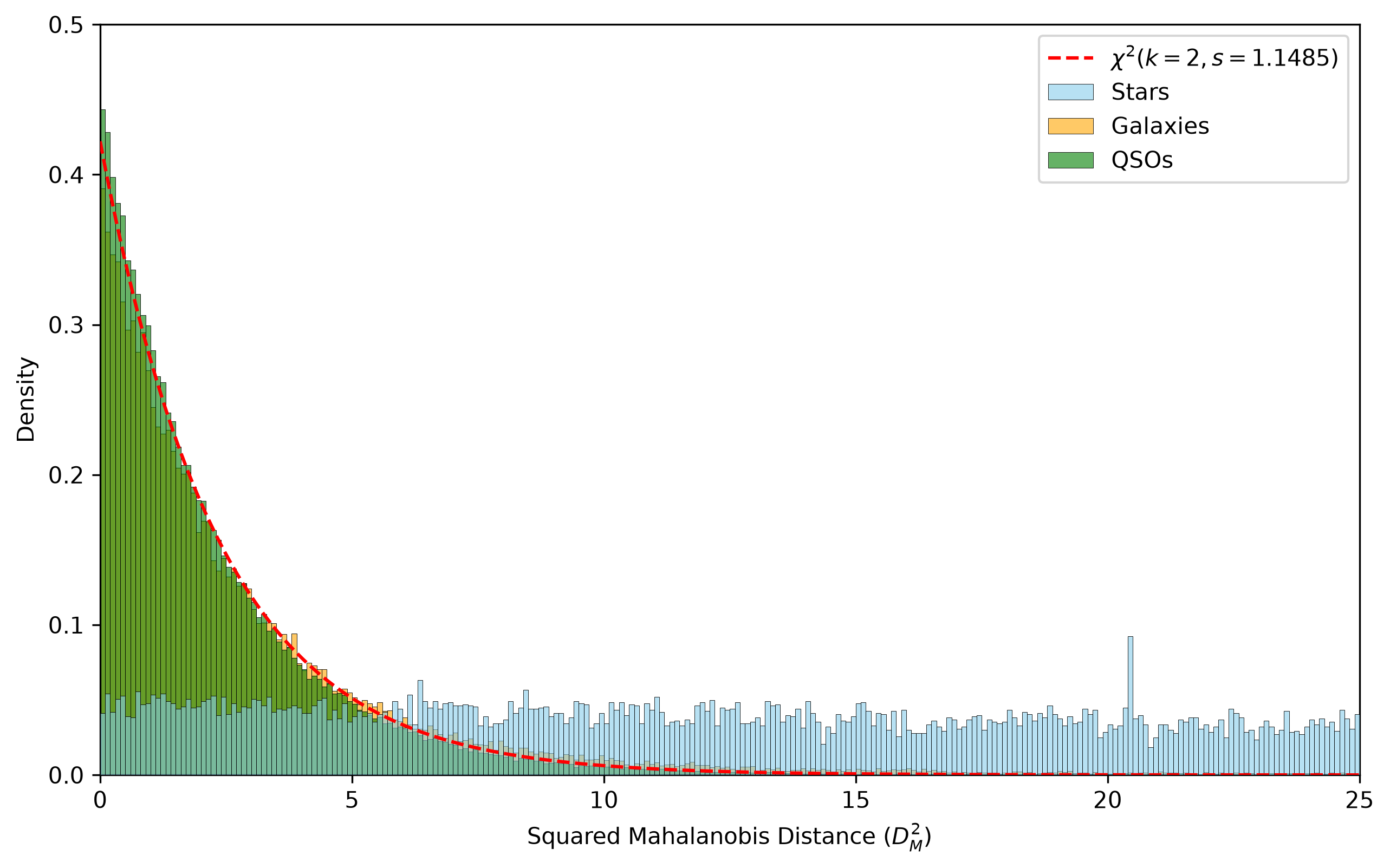}
    \caption{Squared Mahalanobis distance of the proper motion for the quasar, galaxy, and star samples.}
    \label{fig:m_dist_sq}
\end{figure}

We measured the squared Mahalanobis distance for these three samples and obtained the distributions shown in Fig.~\ref{fig:m_dist_sq}. This figure shows that quasars and galaxies, being distant sources, follow the $\chi^2(k=2,s=1.1485)$ distribution, whereas the stellar sample is distributed roughly uniformly along the x-axis. 

Clearly, this criterion does not separate galaxies from quasars. The next criterion is therefore required.

\subsection{\texttt{ruwe} in the Gaia DR3 catalogue}
\label{RUWE}

The \textit{Gaia} DR3 catalogue provides the \texttt{ruwe} as a crucial indicator of the quality of astrometric solutions. This statistic is mathematically defined as
\begin{equation}
    \text{ruwe} = \frac{\sqrt{ \chi^2_{\text{AL}} / (N_{\text{good,AL}} - m) }}{f(G, G_{\text{BP}-\text{RP}})},
    \label{eq:ruwe}
\end{equation}
where $\chi^2_{\text{AL}}$ represents the astrometric goodness-of-fit (\texttt{astrometric\_chi2\_al}) in the along-scan (AL) direction, and $N_{\text{good,AL}}$ is the number of good AL observations (\texttt{astrometric\_n\_good\_obs\_al}). The variable $m$ denotes the number of solved astrometric parameters, which was derived from the number of set bits in the \texttt{astrometric\_params\_solved} parameter. The denominator, $f(G, G_{\text{BP}-\text{RP}})$, was determined from an off-line statistical analysis of secondary solutions \citep{vanleeuwenGaiaDR3Documentation2022,lindegrenGaiaEarlyData2021}. The parameter $f$ is a statistical correction that we used to mitigate systematic biases associated with the $G$-band magnitude and colour of a source \citep{Lindegren2018RUWE}.

\begin{figure}[ht!]
    \centering
    \includegraphics[width=\hsize]{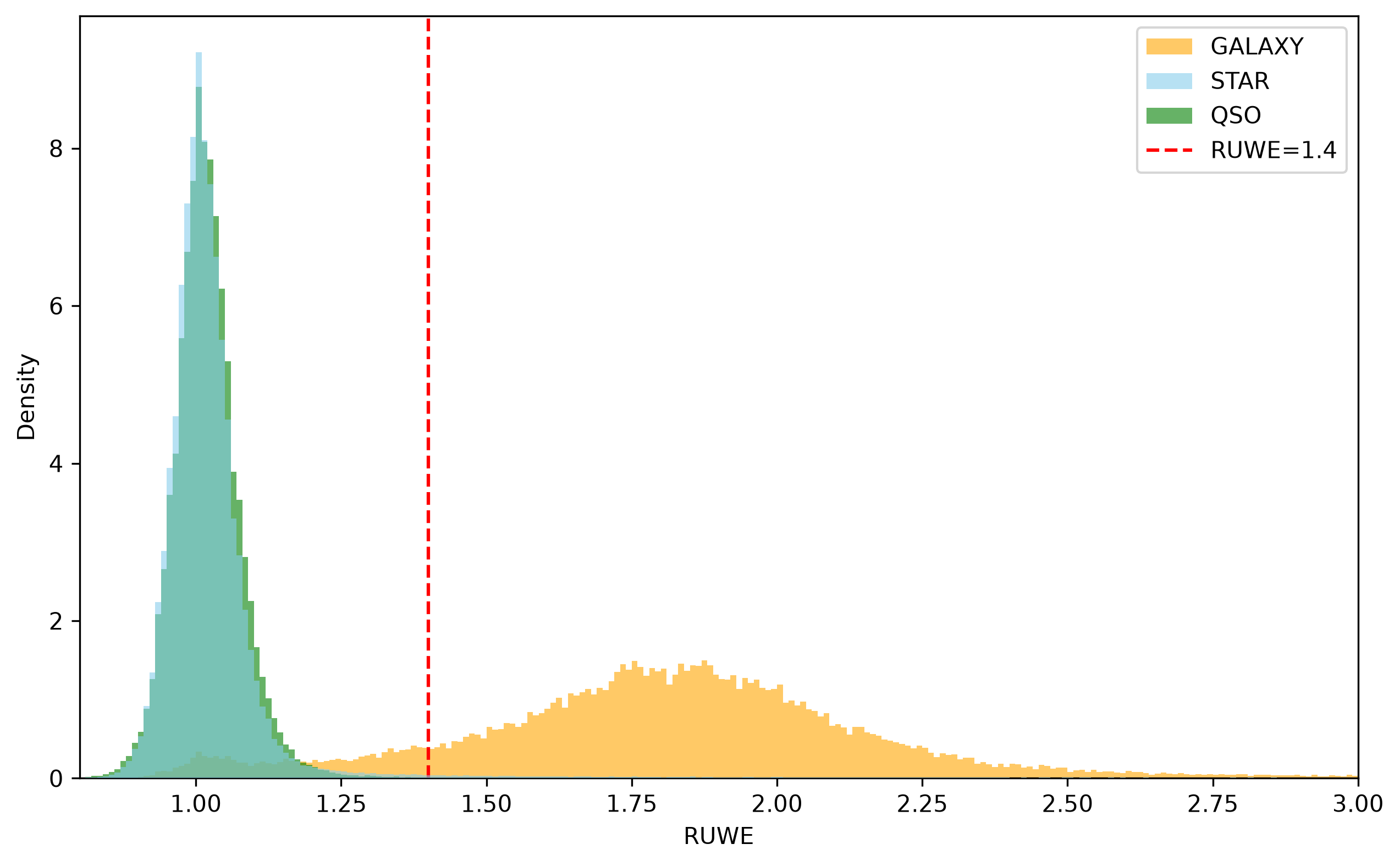}
    \caption{\texttt{ruwe} distributions for the star, galaxy, and quasar samples.}
    \label{fig:ruwe}
\end{figure}

For well-behaved point sources, the \texttt{ruwe} value is theoretically expected to be around 1.0. In contrast, galaxies are spatially extended sources, which means that their astrometric fits in Gaia are typically poorer than those of point-like quasars, resulting in elevated \texttt{ruwe} values. Consequently, the \texttt{ruwe} parameter serves as an effective discriminant \citep{gaiacollaborationGaiaDataRelease2023a}. Figure~\ref{fig:ruwe} illustrates the \texttt{RUWE} distributions for the quasar, galaxy, and stellar reference samples, consistent with the samples presented in Fig.~\ref{fig:m_dist_sq}.

To evaluate the effect of \texttt{ruwe} on our selection, we used samples of stars, galaxies, and quasars. The threshold \texttt{ruwe} \(< 1.4\) is widely adopted in the literature, for example by \citet{fabriciusGaiaEarlyData2021}, to remove potential non-single objects. The comparison results are shown in Table~\ref{tab:ruwe}.
Ultimately, we adopted the condition $\mathrm{ruwe} < 1.4$ to strike an optimal balance between filtering out Galactic contaminants and retaining a maximum number of quasar candidates, as demonstrated in Table~\ref{tab:ruwe}.

\begin{table}[ht!]
    \centering
    \caption{Comparison of the fractions of the star, galaxy, and quasar samples satisfying \texttt{ruwe} $< 1.4$}
    \begin{tabular}{l c c}
    \hline\hline
    Sample & \texttt{ruwe} $< 1.4$  \\
    \hline
    Stars (615\,304)    & $96.86\%$ (595\,983)\\
    Galaxies (54\,035)  & $11.58\%$ (6\,259) \\
    Quasars (168\,060)  & $99.79\%$ (167\,708)\\
    \hline
    \end{tabular}
    \label{tab:ruwe}
\end{table}

\section{Selection result and validation}
\label{sec:res}

Based on the preceding discussion, our updated selection criteria are summarised as follows:
\begin{itemize}
    \item \(\left|\frac{\varpi+0.017\,\mathrm{mas}}{\sigma_{\varpi}}\right|<2 \)
    \item $D^2_{M(\mu)}<7.097$
    \item $\mathrm{ruwe}<1.4$ 
\end{itemize}
The corresponding ADQL query is
{\small
\begin{verbatim}
SELECT * FROM gaiadr3.gaia_source AS g 
WHERE abs((g.parallax + 0.017)/g.parallax_error) < 2 
  AND (power(g.pmra/g.pmra_error,2)
  +power(g.pmdec/g.pmdec_error,2) 
  -2*g.pmra_pmdec_corr*
  g.pmra/g.pmra_error*
  g.pmdec/g.pmdec_error)/
  (1-power(g.pmra_pmdec_corr,2)) < 7.097
  AND g.ruwe<1.4
\end{verbatim}}

To validate the improved performance of our updated selection criteria, we used the SDSS DR16 quasar catalogue (DR16Q) as a reference sample. We cross-matched the DR16Q catalogue with Gaia DR3 using a $1\arcsec$ matching radius, selecting all visually confirmed quasars (including standard quasars, BAL quasars, and blazars), yielding a sample of 168\,060 quasars. By evaluating the recovery rate of these confirmed quasars for both selection methods, we demonstrate the efficiency of our new method. The results are summarised in Table~\ref{tab:sdss_dr16q_comp} and show that the new selection performs better.

\begin{table}[ht!]
    \centering
    \caption{Comparison of the quasar completeness rates using the SDSS DR16Q reference catalogue. The results indicate that the updated selection method achieves a significantly higher completeness for known quasars.}
    \begin{tabular}{c c c}
    \hline\hline
    Sample & Old Selection & New Selection \\
    \hline
    All (168\,060) & $79.00\%$ (132\,760) & $90.25\%$ (151\,671) \\
    \hline
    \end{tabular}
    \label{tab:sdss_dr16q_comp}
\end{table}

\section{Application: Identification of new very luminous $z>2$ quasars}
\label{sec:application}

As an application of our proposed selection schema, we set up a search for 
previously undiscovered very luminous $z>2$ quasars in a particular part of the northern sky, with 
right ascension between 90 and 240 degrees and declination higher than 0. 
Luminous quasars are interesting as targets for high-resolution spectroscopy, 
for instance for abundance measurements or studies of the Ly$\alpha$ forest \citep[e.g.,][]{Schindler2019}.

The initial sample was large, comprising 402\,396 candidates that were identified using astrometric data alone. To focus on our objectives and accelerate our research, we incorporated photometric data from the Two Micron All Sky Survey~(2MASS) and the Wide-Field Infrared Survey Explorer~(WISE) to filter probable quasars. Consequently, we specifically searched for objects that met the following criteria:
\begin{enumerate}
    \item Proper motion and parallax consistent with 0, as described above.
    \item 2MASS $J$-band magnitude brighter than 16.
    \item WISE colour $W1-W2 > 0.4$
    \item No spectroscopic redshift listed in the NASA Extragalactic Database (NED)\footnote{\url{https://ned.ipac.caltech.edu}}.
\end{enumerate}

Criterion 1 is obvious. Criterion 2 was set to select luminous quasars if at $z>2$. Criterion 3 was set to 
remove most of the stellar sources while still maintaining the possibility to select quasars 
in the redshift range where this colour is smallest \citep[see][their Fig. 1, and Table~\ref{tab:spherex_quasars}]{2012ApJ...753...30S}.
Criterion 4 removed most of the quasars with previous spectroscopic confirmation. 

\begin{table*}[ht!]
\small
\renewcommand{\arraystretch}{0.9}
\centering
\caption{SPHEREx-selected quasars.}
\label{tab:spherex_quasars}
\begin{tabular}{lcccccc}
\hline\hline
Target & RA (J2000) & Dec (J2000) &  QSO $A_V$ & $W1-W2$ & Redshift & Other Observations \\
 & \textit{[hr:min:sec]} & \textit{[deg:min:sec]} & \textit{[mag]} & \textit{[mag]} & \\
\hline
Gaia DR3 714636898258763008 & 09:19:42.73 & +35:57:05.19 & 0.034 & 0.98 & $2.3668\pm0.0012$ & $z_{\mathrm{DESI}}=0.1795$, BAL \\
Gaia DR3 761680495389089536 & 11:10:05.93 & +34:28:08.72 & 0.199 & 0.83 & $2.6065\pm0.0015$ & $z_{\mathrm{DESI}}=1.6772$, BAL \\
Gaia DR3 763275719257657088 & 11:04:42.05 & +37:56:14.90 & 0.000 & 0.99 & $2.6845\pm0.0033$ & $z_{\mathrm{LAMOST}}=2.6362$, BAL \\
Gaia DR3 890044221621193984 & 07:12:56.46 & +31:42:21.57 & 0.000 & 1.12 & $2.3382\pm0.0023$ & - \\
Gaia DR3 941441094524786944 & 06:39:02.76 & +34:16:39.07 & 0.209 & 0.98 & $2.9150\pm0.0041$ & - \\
Gaia DR3 948402927272355456 & 07:15:43.13 & +40:32:59.56  & 0.000 & 1.04 & $2.1639\pm0.0013$ & $z_{\mathrm{DESI}}=2.1573$ \\
Gaia DR3 967799308818609280 & 06:33:51.78 & +48:47:23.54 & 0.000 & 1.02 & $2.1428\pm0.0023$ & - \\
Gaia DR3 976118321169640704 & 07:32:38.27 & +49:11:32.19 & 0.000 & 0.82 & $2.7246\pm0.0020$ & - \\
Gaia DR3 999116492355319424 & 06:05:36.32 & +57:47:09.34 & 0.000 & 0.84 & $2.0727\pm0.0016$ & - \\
Gaia DR3 1056363011931511552 & 11:33:55.82 & +65:18:16.45 & 0.264 & 0.70 & $2.7405\pm0.0015$ & $z_{\mathrm{DESI}}=1.2686$, BAL \\
Gaia DR3 1085417851509571456 & 07:38:41.64 & +58:12:47.07 & 0.000 & 0.74 & $2.2146\pm0.0019$ &  $z_{\mathrm{SDSS}}=0.1195$, BAL \\
Gaia DR3 1093392781144614144 & 08:33:28.48 & +65:54:39.95 & 0.000 & 1.23 & $2.0117\pm0.0033$ & $z_{\mathrm{DESI}}=2.0134$, BAL \\
Gaia DR3 1099763076637665536 & 07:08:02.48 & +63:15:59.67 & 0.000 & 0.56 & $3.3394\pm0.0032$ & - \\
Gaia DR3 1148778858327502976 & 07:27:04.64 & +82:32:56.19 & 0.000 & 0.79 & $2.1953\pm0.0015$ & - \\
Gaia DR3 1183404261896916736 & 15:05:18.97 & +13:54:48.16 & 0.000 & 0.81 & $2.3614\pm0.0007$ &  $z_{\mathrm{DESI}}=0.3732$, BAL \\
Gaia DR3 1231945673038241408 & 14:16:21.34 & +15:25:31.33 & 0.023 & 1.08 & $2.2329\pm0.0017$ & $z_{\mathrm{SDSS}}=1.5615$, BAL \\
Gaia DR3 1268454033221308160 & 15:04:40.56 & +26:54:58.67 & 0.000 & 0.71 & $2.9343\pm0.0021$ & $z_{\mathrm{LAMOST}}=2.9508$, BAL \\
Gaia DR3 1485032844363848704 & 14:20:18.30 & +39:39:12.78 & 0.010 & 0.95 & $2.6049\pm0.0010$ & \shortstack[l]{$z_{\mathrm{LAMOST}}=2.5900$\\ $z_{\mathrm{DESI}}=2.5418$} \\
Gaia DR3 1523063611538852992 & 13:10:38.57 & +38:03:33.46 & 0.000 & 1.08 & $2.5852\pm0.0016$ & $z_{\mathrm{DESI}}=2.5453$ \\
Gaia DR3 1673295859540228864 & 14:20:05.12 & +68:10:47.29 & 0.334 & 0.89 & $3.2016\pm0.0013$ & - \\
Gaia DR3 3157390438235296000 & 06:52:20.86 & +08:26:01.88 & 0.000 & 0.49 & $2.1840\pm0.0071$ & - \\
Gaia DR3 3157933322101067904 & 07:00:55.75 & +09:46:20.16 & 0.000 & 0.95 & $2.3850\pm0.0021$ & $z_{\mathrm{LAMOST}}=1.7190$, BAL \\
Gaia DR3 3365048499220959488 & 06:59:24.34 & +19:11:44.56 & 0.000 & 1.18 & $2.0208\pm0.0024$ & - \\
Gaia DR3 3986496938005599232 & 10:49:34.90 & +19:35:43.18 & 0.000 & 0.75 & $2.6488\pm0.0030$ &  $z_{\mathrm{DESI}}=2.6489$ \\
\hline
\end{tabular}
\end{table*}

\begin{figure}[ht!]
\centering
\includegraphics[width=0.48\textwidth]{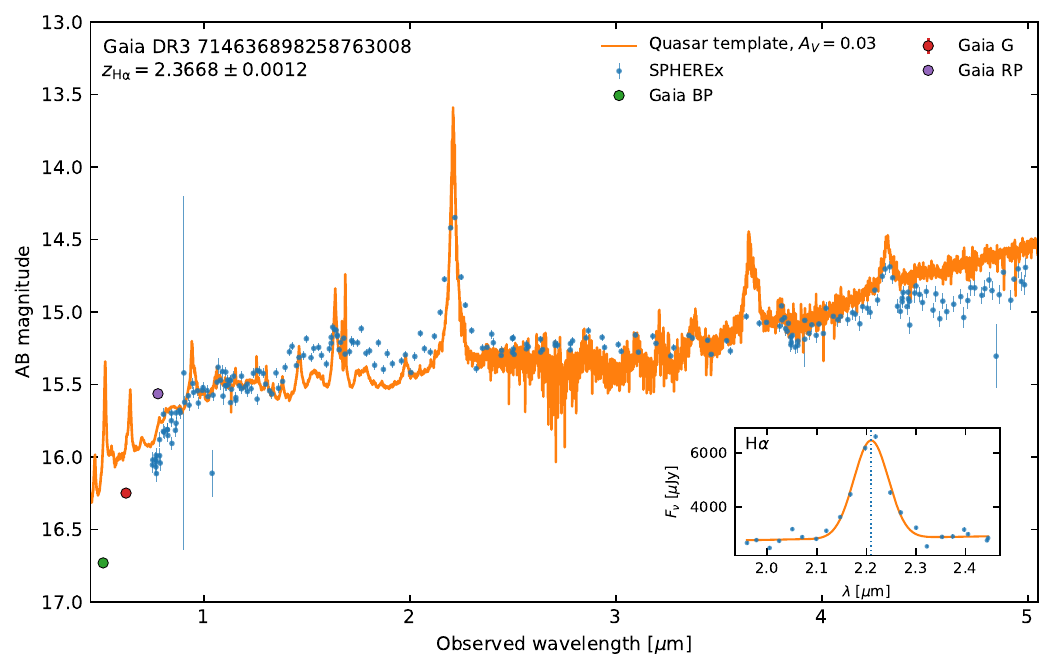}
\caption{Spectrum (AB magnitude vs. observed wavelength in 
$\mu m$) of the source Gaia DR3 714636898258763008. The
SPHEREx spectrum is shown with blue points, and the Gaia photometry is
also indicated with green, red, and purple points. We overplot the quasar
template from \citet{Selsing2016} at the redshift determined from 
fitting the H$\alpha$ emission line. The determined redshift is provided
in the upper left corner of the figure and in Table~\ref{tab:spherex_quasars}. The inset in the lower 
right corner shows the fit to the H$\alpha$ line. The quasar template is
reddened using the SMC extinction curve from \citet{1992ApJ...395..130P} 
after correction for Galactic extinction. The figures for the other 23 candidates have the same structure and can be found in Appendix~\ref{spec}. 
}
\label{fig:714636898258763008}
\end{figure}

This selection left us with about 600 sources. For all of these, we downloaded SPHEREx spectra that covered
the wavelength range from about 0.7 to 5 $\mu m$ \citep{2026ApJ...999..139B,2026arXiv260310135D}. We visually inspected all of these spectra. About two-thirds are quasars and the rest are galaxies, typically cuspy elliptical 
galaxies at redshifts below a few tenths. Quasars are easy to identify from the strong H$\alpha$ emission
line. In Table~\ref{tab:spherex_quasars} we provide details for all 24 $z>2$ quasars we identified,
and in Fig.~\ref{fig:714636898258763008} and Appendix~\ref{spec}, we show the spectrum for each 
source together with the Gaia photometry. For each source, we corrected for Galactic extinction, using the 
extinction maps from \citet{2011ApJ...737..103S}, and determined 
the redshift by fitting a Gaussian to the H$\alpha$ emission line (the fit to the H$\alpha$ line is 
shown with an inset for each quasar). We also estimated the intrinsic extinction in the QSO by fitting the 
spectrum with the quasar template from \citet{Selsing2016}. We assumed that the dust is at the redshift 
of the quasar and that the extinction curve is similar to that of the Small Magellanic Cloud (SMC), as 
quantified by \citet{1992ApJ...395..130P}. In addition, seven targets are also listed in \citet{2026A&A...708A.363P}.

We also searched the SDSS, LAMOST, and DESI catalogues and found several relatively recent spectra that had not yet been included in NED. Nevertheless, some catalogue entries probably report incorrect redshifts. For instance, in the DESI catalogue, Gaia DR3 sources Gaia DR3 714636898258763008, Gaia DR3 761680495389089536, Gaia DR3 1056363011931511552, and Gaia DR3 1183404261896916736 are misclassified as galaxies with incorrect redshifts. Moreover, Gaia DR3 1056363011931511552 is assigned a redshift that is lower than the redshift we inferred from
H$\alpha$. Gaia DR3 3157933322101067904 is also assigned an incorrect redshift in the LAMOST catalogue. These targets are BAL quasars, which can cause automated classification pipelines to become incorrect.

\section{Discussion and conclusions}
\label{sec:dis}

We have presented a strategy for using Gaia DR3 to select quasars purely based on astrometry.
In particular, we showed how a 2$\sigma$ cut on proper motion can be implemented, for example, and how the correlation between the measurements of proper motion in $\alpha$ and $\delta$ can be taken into account.

We note that the \texttt{ruwe} criterion introduced in Sect.~\ref{RUWE} is important for filtering out extended sources such as galaxies and retaining point sources. However, in some cases, researchers may wish to study lower-luminosity active galactic nuclei~(AGNs), for which the host can be of comparable brightness as the AGN. In this case, the criterion $\mathrm{ruwe}<1.4$ should be removed. 

We quantified how our strategy performs compared to previous strategies in the literature by studying the astrometric properties of objects in SDSS DR16Q. These results indicate that the new strategy offers improved performance in the coverage of previously confirmed quasars.

Finally, we showed an application of our selection strategy in which we carried out a search for new luminous $z>2$ quasars in the northern sky. We identified 24 bright quasars in the redshift range $2.0 < z < 3.3$. These sources are all very interesting targets for follow-up higher-resolution spectroscopy. 

We showed that our selection strategy is viable and performs well. This strategy also provides a potential means of eliminating some highly unreliable candidates from the Gaia QSO candidate catalogue. In particular, it might be used to build a purer quasar sample for training the discrete source classifier (DSC) module in the Gaia pipeline. Such a sample would be less strongly contaminated by misclassified sources.

However, this selection strategy also has limitations. Its efficiency depends strongly on the uncertainty of the astrometric measurements; in other words, the method performs better when the astrometric uncertainties are small. This limitation is expected to be alleviated by future Gaia data releases, such as DR4 and DR5, which will provide more accurate astrometry and thus enable the construction of a more reliable quasar sample.

Beyond measurement uncertainties, other effects might also affect the astrometric properties of quasars. For example, \citet{gaiacollaborationGaiaEarlyData2021} found a systematic proper motion of $\sim5\mu \mathrm{as}/\mathrm{yr}$ for background objects such as quasars due to the acceleration of the Solar System. In addition, unresolved double or multiple AGNs might cause some quasars to appear to have proper motion in the optical band \citep{makarovQuasarsProperMotions2022}; quasar variability is another potential reason why they might be measured with proper motion \citep{bachchanGaiaReferenceFrame2016}.

We finally note that there is substantial potential for citizen science in the combination of
Gaia-based selection of quasar candidates and spectroscopic verification using SPHEREx data. It would 
be relatively easy also for the interested citizen to identify interesting quasars.

\begin{acknowledgements}
This work has made use of data from the European Space Agency (ESA) mission Gaia, processed by the Gaia Data Processing and Analysis Consortium (DPAC). Funding for the DPAC has been provided by national institutions, in particular the institutions participating in the Gaia Multilateral Agreement.
This publication makes use of data products from the Spectro-Photometer for the History of the Universe, Epoch of Reionization and Ices Explorer (SPHEREx), which is a joint project of the Jet Propulsion Laboratory and the California Institute of Technology, and is funded by the National Aeronautics and Space Administration.
This research has made use of the NASA/IPAC Extragalactic Database, which is funded by the National Aeronautics and Space Administration and operated by the California Institute of Technology.
Funding for the Sloan Digital Sky Survey IV has been provided by the Alfred P. Sloan Foundation, the U.S. Department of Energy Office of Science, and the Participating Institutions. SDSS acknowledges support and resources from the Center for High-Performance Computing at the University of Utah. The SDSS web site is www.sdss4.org.
SDSS is managed by the Astrophysical Research Consortium for the Participating Institutions of the SDSS Collaboration including the Brazilian Participation Group, the Carnegie Institution for Science, Carnegie Mellon University, Center for Astrophysics | Harvard \& Smithsonian (CfA), the Chilean Participation Group, the French Participation Group, Instituto de Astrofísica de Canarias, The Johns Hopkins University, Kavli Institute for the Physics and Mathematics of the Universe (IPMU) / University of Tokyo, the Korean Participation Group, Lawrence Berkeley National Laboratory, Leibniz Institut für Astrophysik Potsdam (AIP), Max-Planck-Institut für Astronomie (MPIA Heidelberg), Max-Planck-Institut für Astrophysik (MPA Garching), Max-Planck-Institut für Extraterrestrische Physik (MPE), National Astronomical Observatories of China, New Mexico State University, New York University, University of Notre Dame, Observatório Nacional / MCTI, The Ohio State University, Pennsylvania State University, Shanghai Astronomical Observatory, United Kingdom Participation Group, Universidad Nacional Autónoma de México, University of Arizona, University of Colorado Boulder, University of Oxford, University of Portsmouth, University of Utah, University of Virginia, University of Washington, University of Wisconsin, Vanderbilt University, and Yale University.
Guoshoujing Telescope (the Large Sky Area Multi-Object Fiber Spectroscopic Telescope LAMOST) is a National Major Scientific Project built by the Chinese Academy of Sciences. Funding for the project has been provided by the National Development and Reform Commission. LAMOST is operated and managed by the National Astronomical Observatories, Chinese Academy of Sciences.
This research used data obtained with the Dark Energy Spectroscopic Instrument (DESI). DESI construction and operations is managed by the Lawrence Berkeley National Laboratory. This material is based upon work supported by the U.S. Department of Energy, Office of Science, Office of High-Energy Physics, under Contract No. DE–AC02–05CH11231, and by the National Energy Research Scientific Computing Center, a DOE Office of Science User Facility under the same contract. Additional support for DESI was provided by the U.S. National Science Foundation (NSF), Division of Astronomical Sciences under Contract No. AST-0950945 to the NSF’s National Optical-Infrared Astronomy Research Laboratory; the Science and Technology Facilities Council of the United Kingdom; the Gordon and Betty Moore Foundation; the Heising-Simons Foundation; the French Alternative Energies and Atomic Energy Commission (CEA); the National Council of Humanities, Science and Technology of Mexico (CONAHCYT); the Ministry of Science and Innovation of Spain (MICINN), and by the DESI Member Institutions: www.desi.lbl.gov/collaborating-institutions. The DESI collaboration is honored to be permitted to conduct scientific research on I’oligam Du’ag (Kitt Peak), a mountain with particular significance to the Tohono O’odham Nation. Any opinions, findings, and conclusions or recommendations expressed in this material are those of the author(s) and do not necessarily reflect the views of the U.S. National Science Foundation, the U.S. Department of Energy, or any of the listed funding agencies.
This work made use of Astropy\footnote{\tiny{\url{https://www.astropy.org}}}: a community-developed core Python package and an ecosystem of tools and resources for astronomy \citep{astropy:2013, astropy:2018, astropy:2022}.
The Cosmic Dawn Center (DAWN) is funded by the Danish National Research Foundation under grant DNRF140.  
JPUF is supported by the Independent Research Fund Denmark (DFF-4090-00079) and thanks the Carlsberg Foundation for support. JPUF acknowledges support by the European Research Council (ERC) under the European Union’s research and innovation program (ERC Grant HEAVYMETAL No.~101071865). 
      
\end{acknowledgements}

\bibliographystyle{aa}
\bibliography{ref}

\begin{appendix}
\section{SPHEREx spectra of new quasars}
\label{spec}

\begin{figure}[ht!]
\centering
\includegraphics[width=0.48\textwidth]{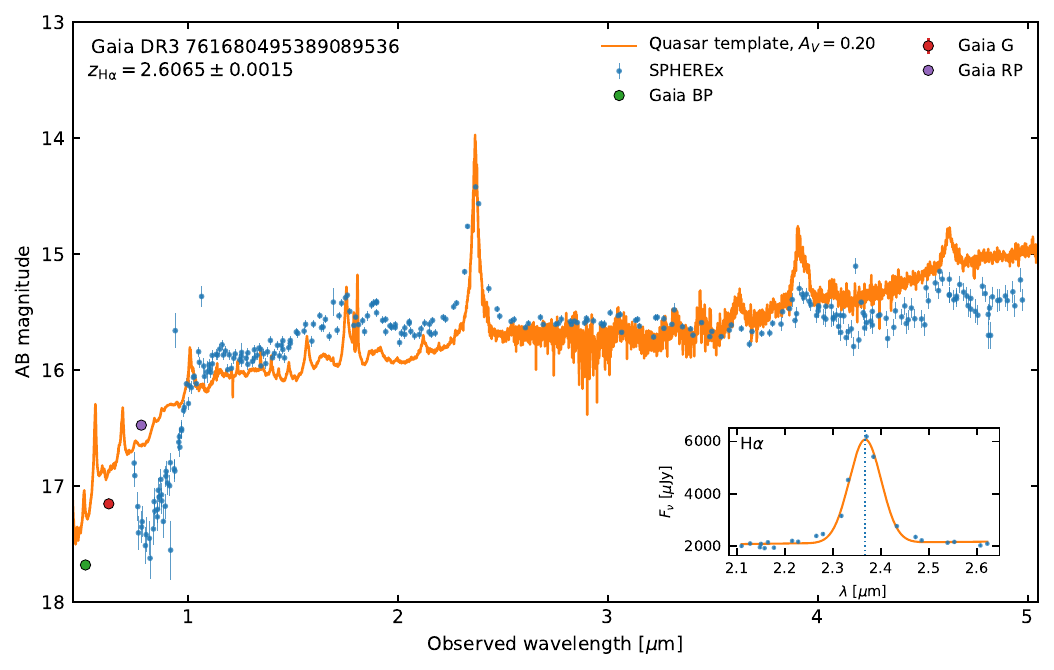}
\caption{Gaia DR3 761680495389089536}
\label{fig:761680495389089536}
\end{figure}

\begin{figure}[ht!]
\centering
\includegraphics[width=0.48\textwidth]{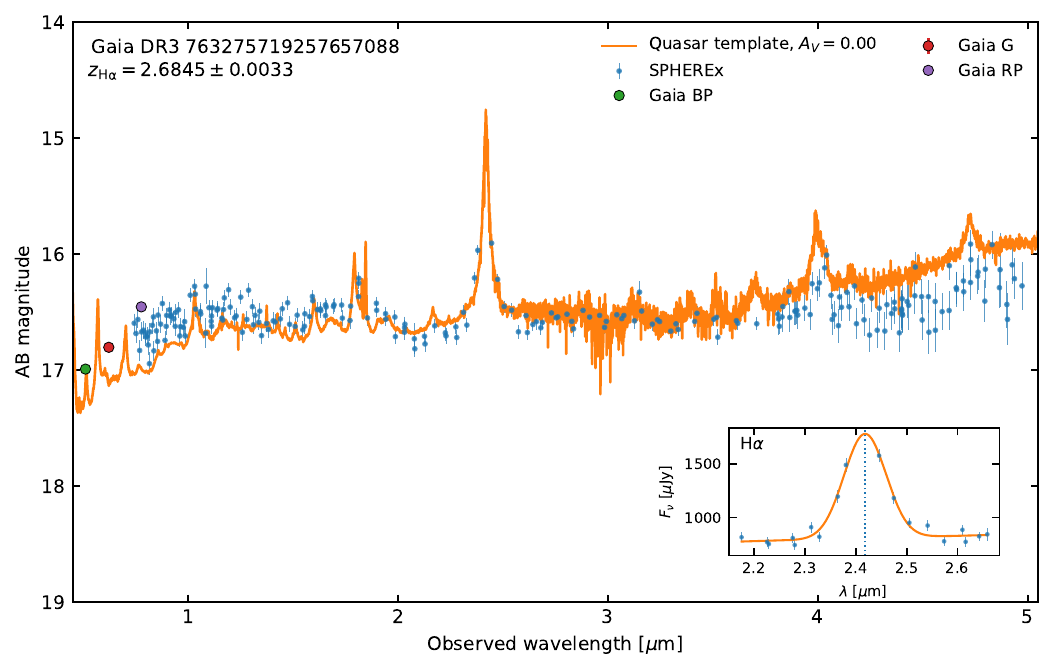}
\caption{Gaia DR3 763275719257657088}
\label{fig:763275719257657088}
\end{figure}

\begin{figure}[ht!]
\centering
\includegraphics[width=0.48\textwidth]{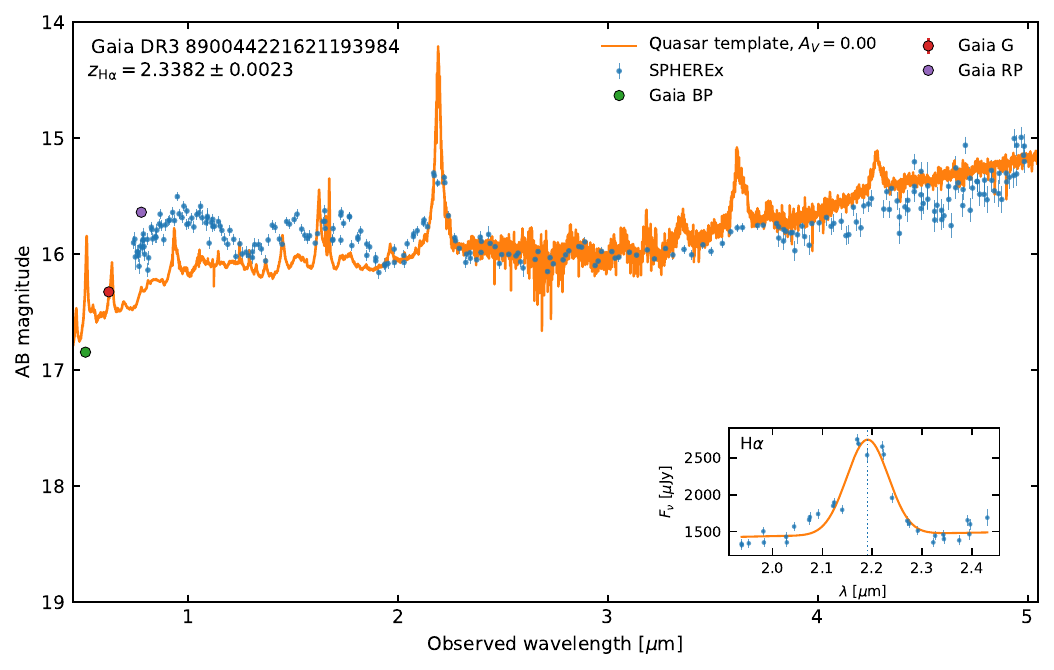}
\caption{Gaia DR3 890044221621193984}
\label{fig:890044221621193984}
\end{figure}

\begin{figure}[ht!]
\centering
\includegraphics[width=0.48\textwidth]{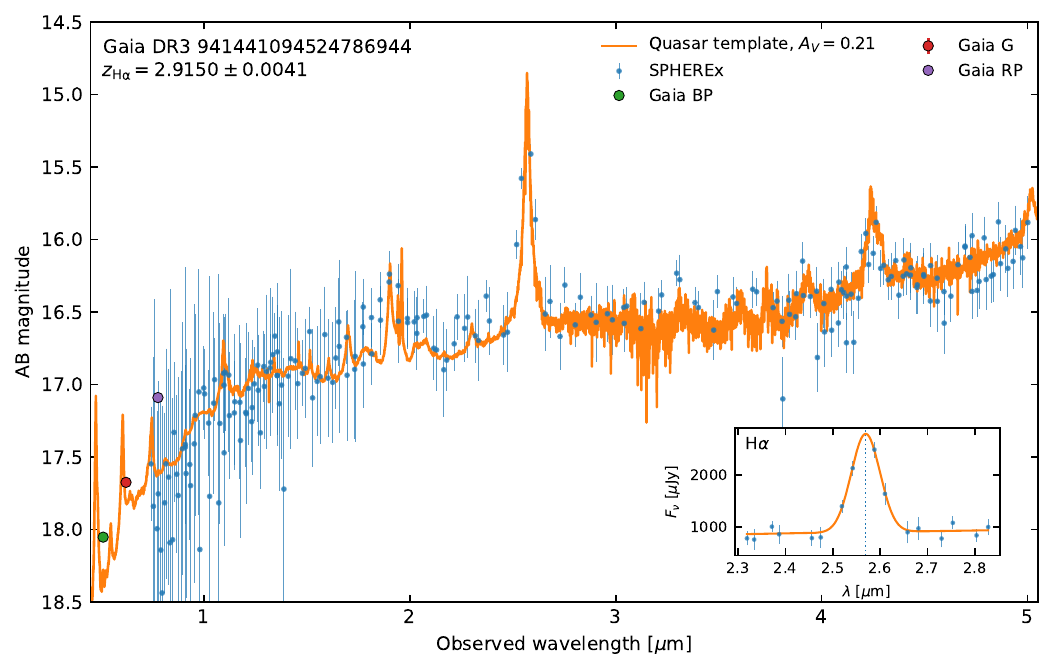}
\caption{Gaia DR3 941441094524786944}
\label{fig:941441094524786944}
\end{figure}

\begin{figure}[ht!]
\centering
\includegraphics[width=0.48\textwidth]{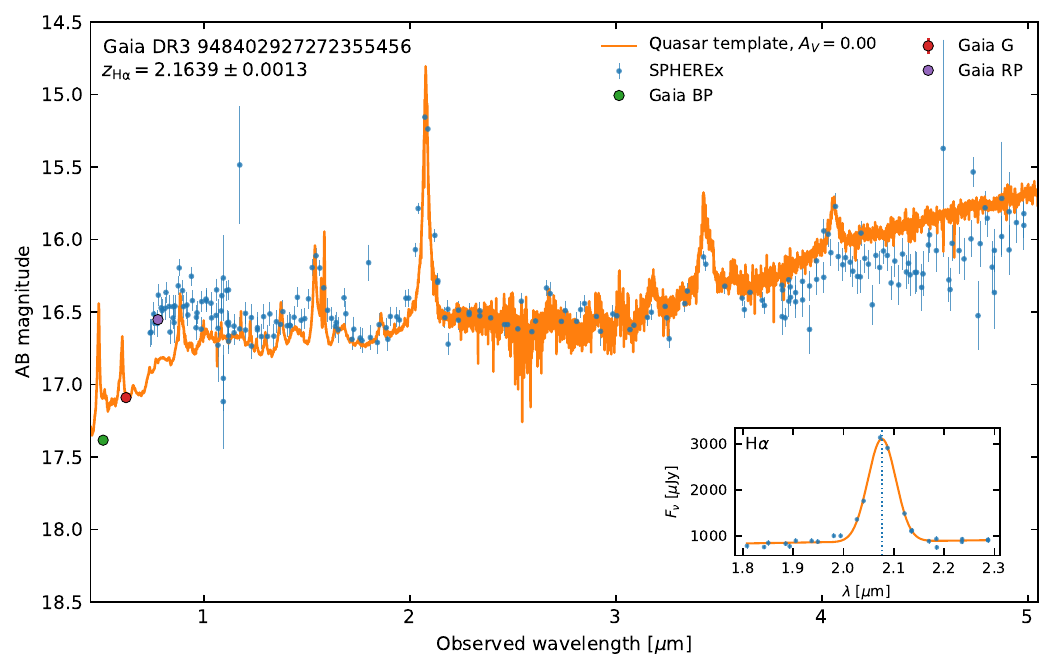}
\caption{Gaia DR3 948402927272355456}
\label{fig:948402927272355456}
\end{figure}

\begin{figure}[ht!]
\centering
\includegraphics[width=0.48\textwidth]{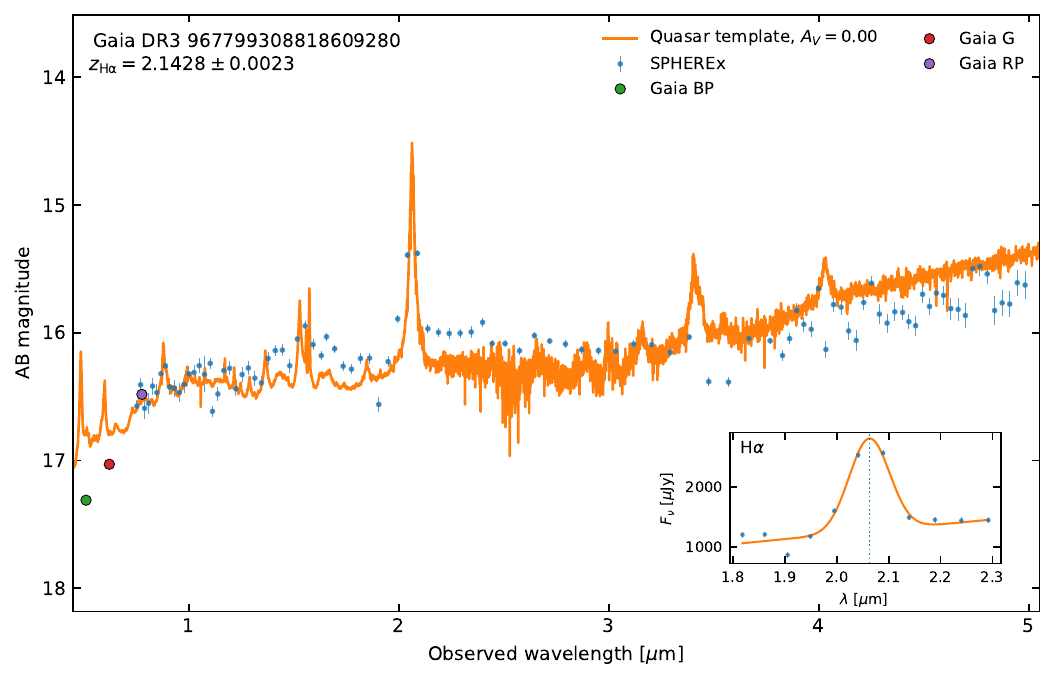}
\caption{Gaia DR3 967799308818609280}
\label{fig:967799308818609280}
\end{figure}

\begin{figure}[ht!]
\centering
\includegraphics[width=0.48\textwidth]{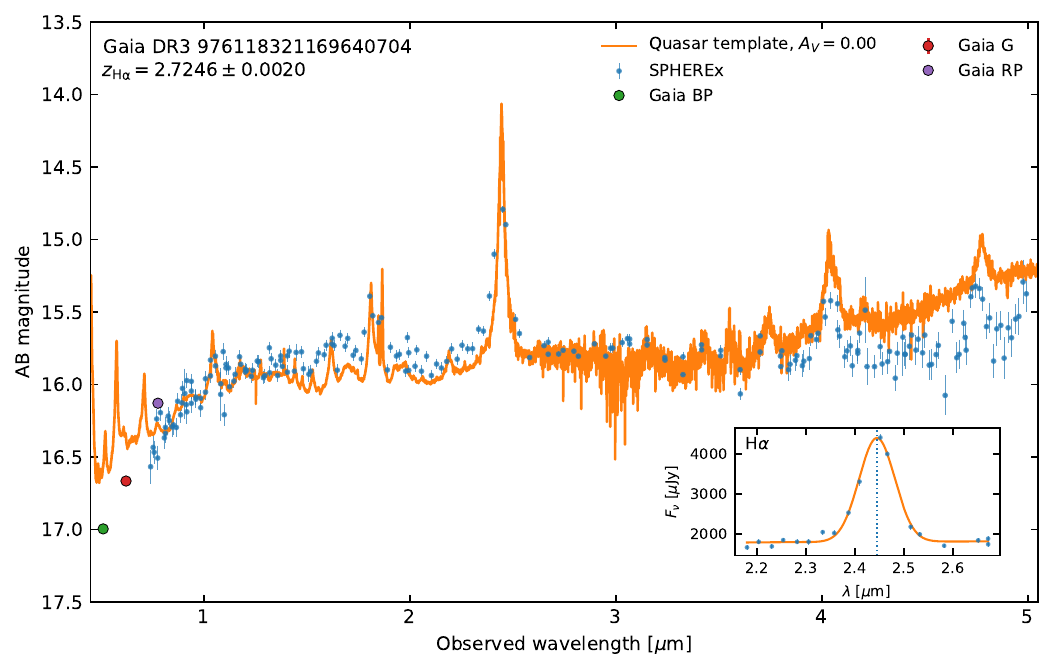}
\caption{Gaia DR3 976118321169640704}
\label{fig:976118321169640704}
\end{figure}

\begin{figure}[ht!]
\centering
\includegraphics[width=0.48\textwidth]{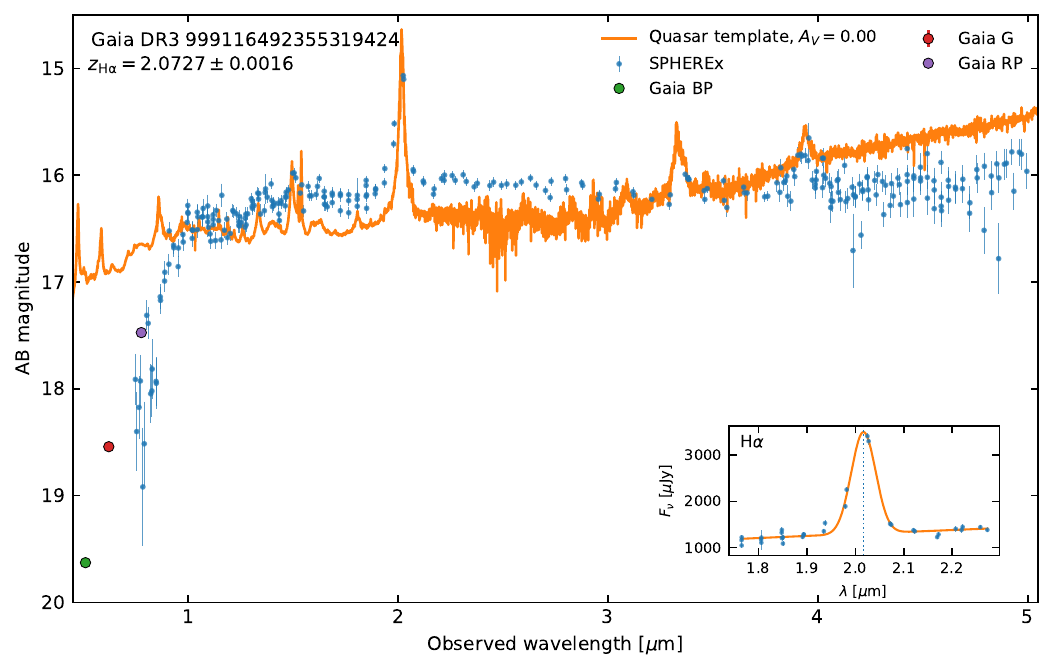}
\caption{Gaia DR3 999116492355319424}
\label{fig:999116492355319424}
\end{figure}

\begin{figure}[ht!]
\centering
\includegraphics[width=0.48\textwidth]{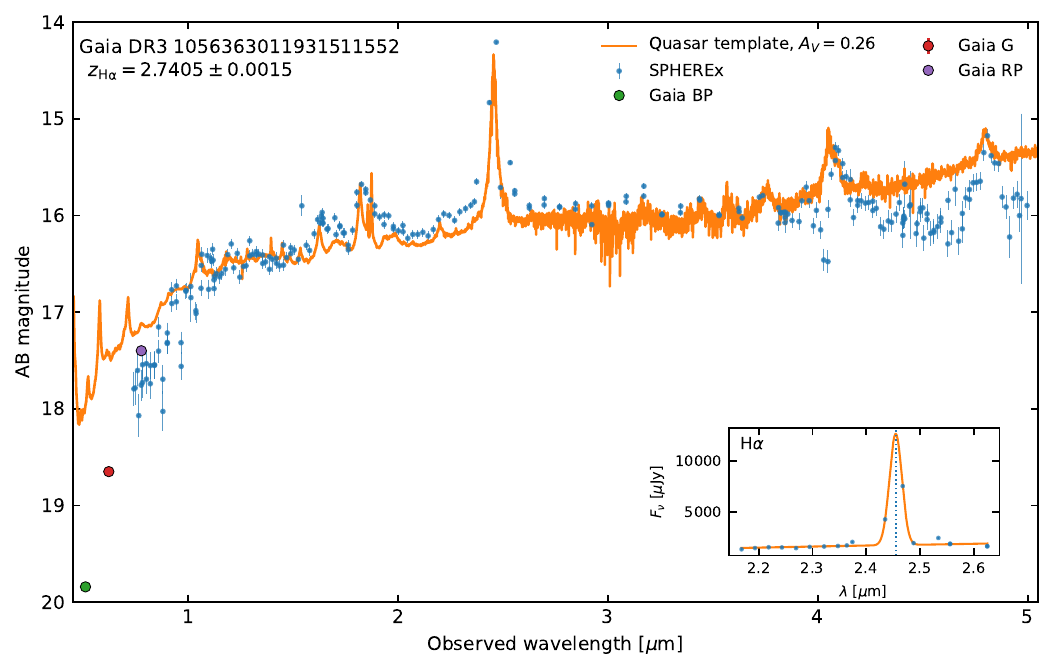}
\caption{Gaia DR3 1056363011931511552}
\label{fig:1056363011931511552}
\end{figure}

\begin{figure}[ht!]
\centering
\includegraphics[width=0.48\textwidth]{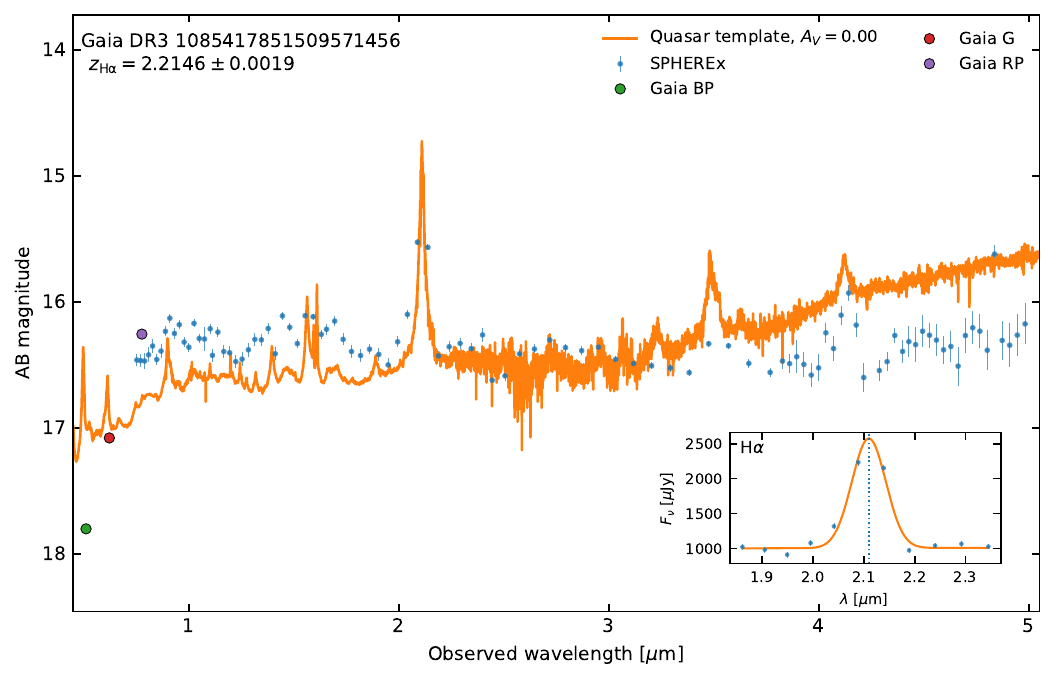}
\caption{Gaia DR3 1085417851509571456}
\label{fig:1085417851509571456}
\end{figure}

\begin{figure}[ht!]
\centering
\includegraphics[width=0.48\textwidth]{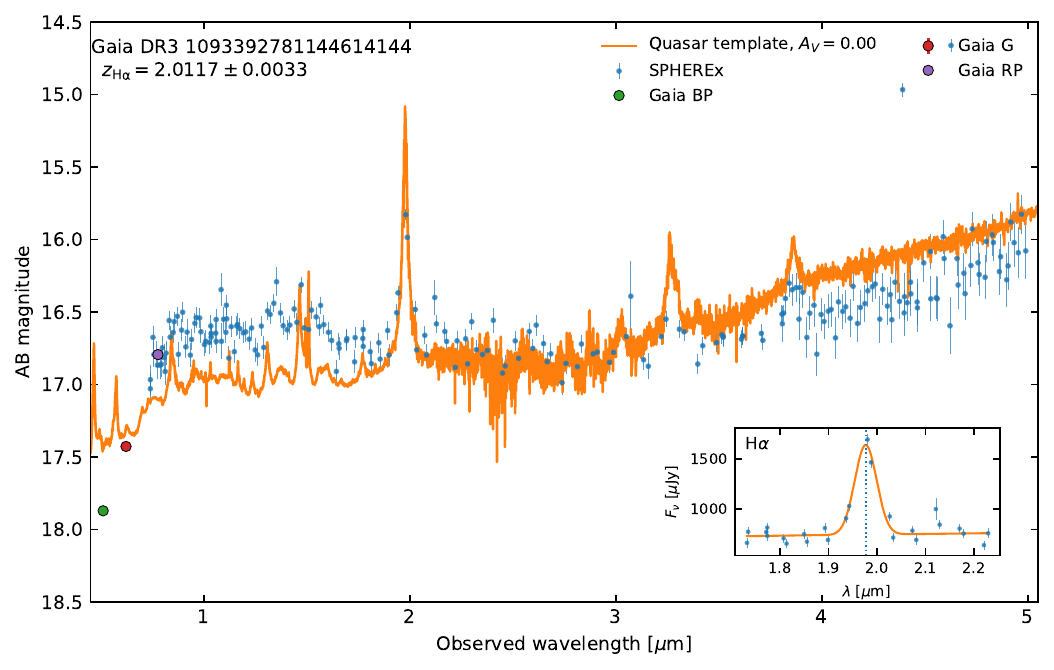}
\caption{Gaia DR3 1093392781144614144}
\label{fig:1093392781144614144}
\end{figure}

\begin{figure}[ht!]
\centering
\includegraphics[width=0.48\textwidth]{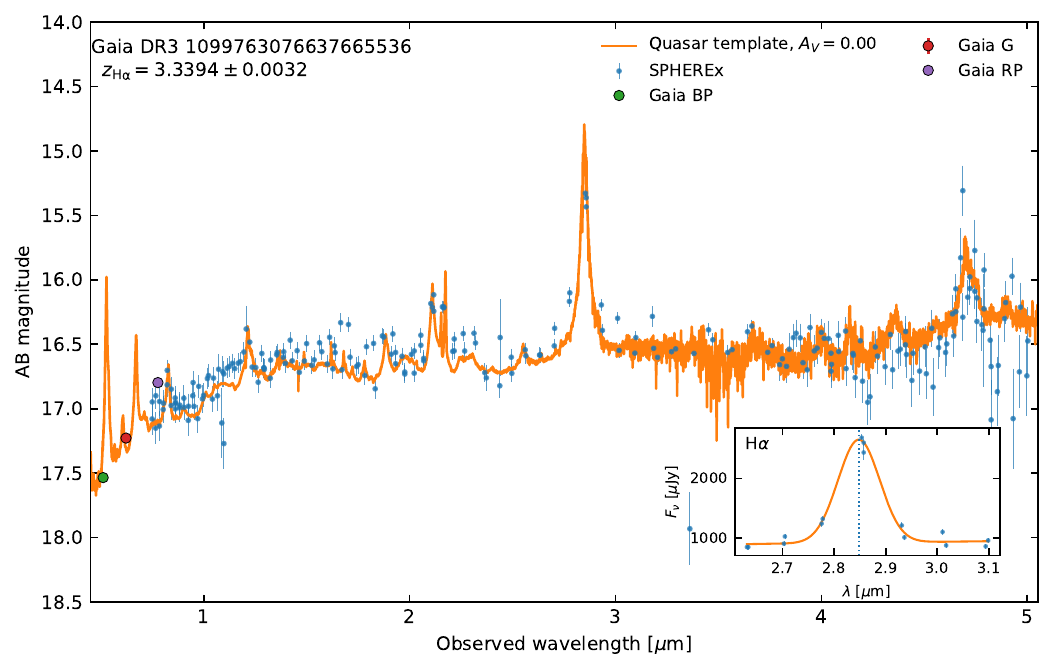}
\caption{Gaia DR3 1099763076637665536}
\label{fig:1099763076637665536}
\end{figure}

\begin{figure}[ht!]
\centering
\includegraphics[width=0.48\textwidth]{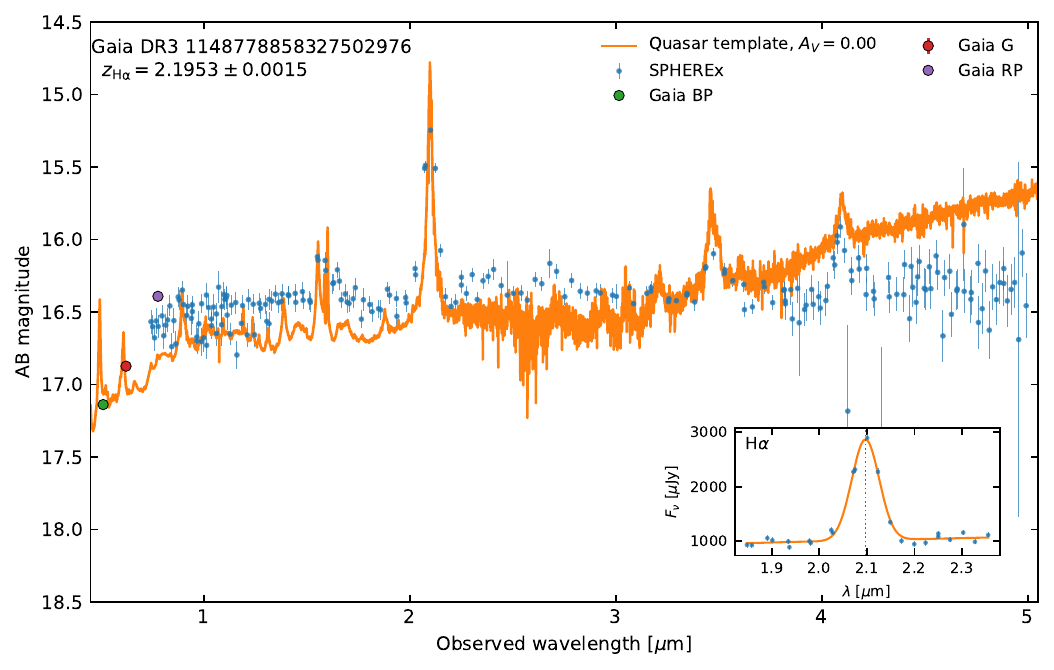}
\caption{Gaia DR3 1148778858327502976}
\label{fig:1148778858327502976}
\end{figure}

\begin{figure}[ht!]
\centering
\includegraphics[width=0.48\textwidth]{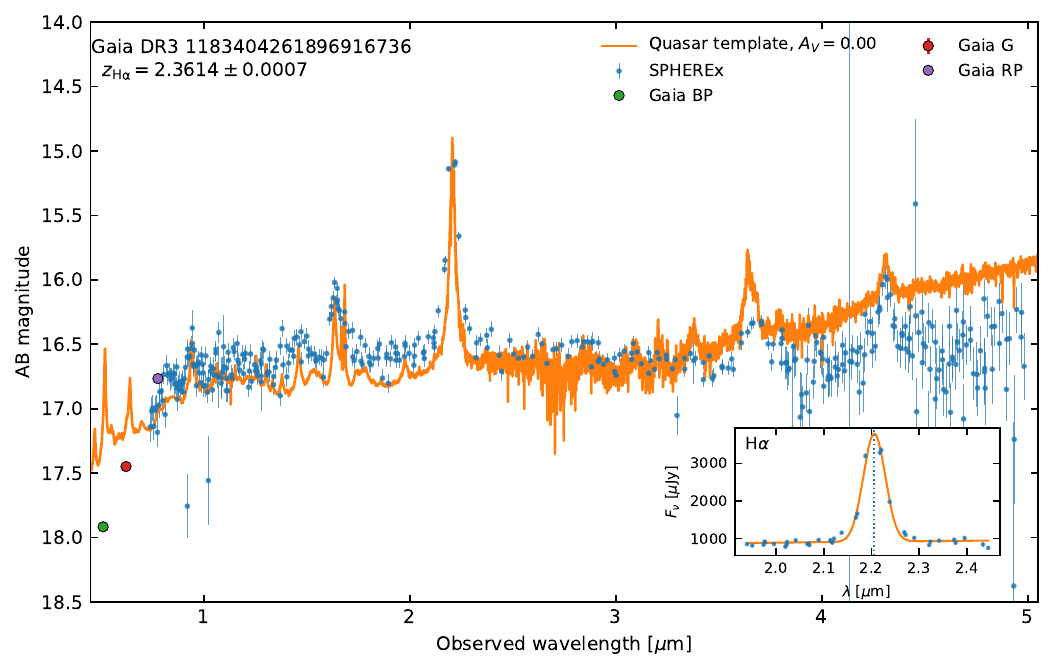}
\caption{Gaia DR3 1183404261896916736}
\label{fig:1183404261896916736}
\end{figure}

\begin{figure}[ht!]
\centering
\includegraphics[width=0.48\textwidth]{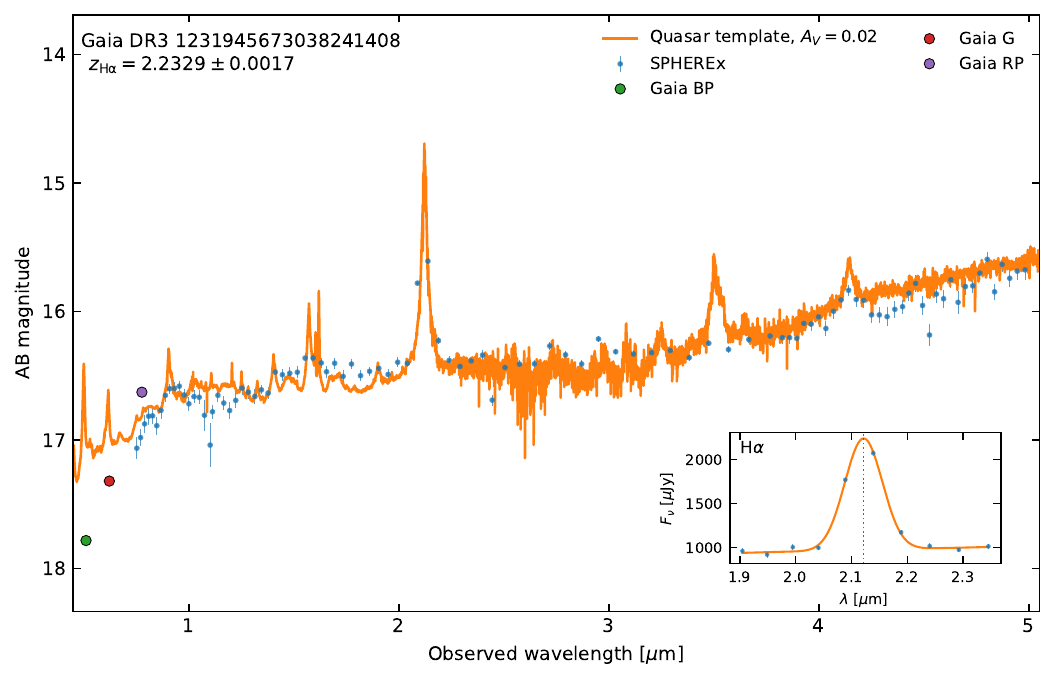}
\caption{Gaia DR3 1231945673038241408}
\label{fig:1231945673038241408}
\end{figure}

\begin{figure}[ht!]
\centering
\includegraphics[width=0.48\textwidth]{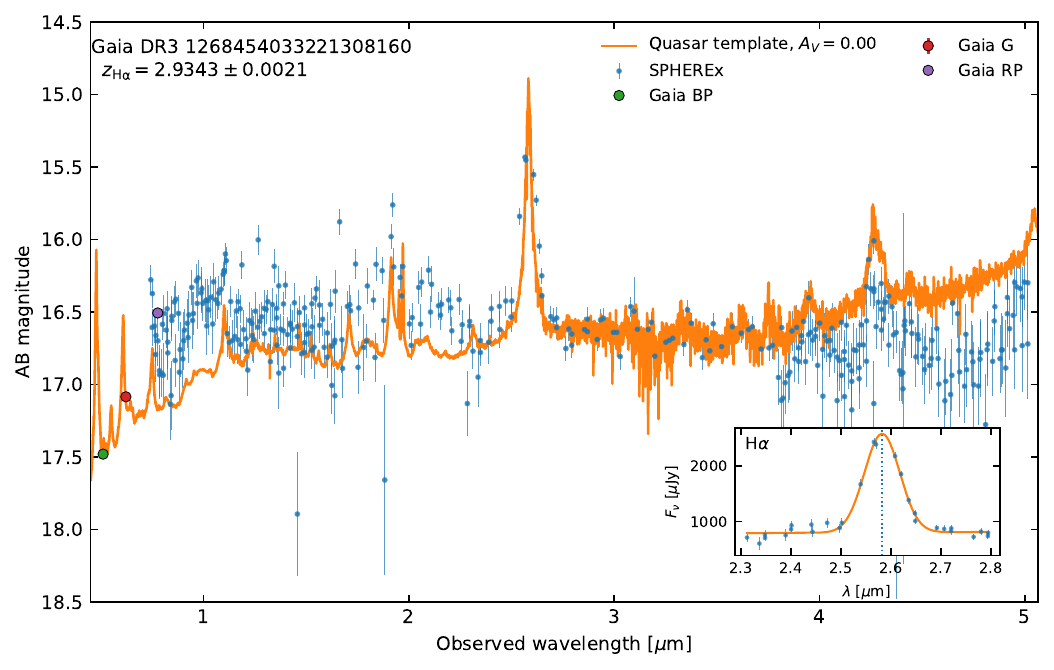}
\caption{Gaia DR3 1268454033221308160}
\label{fig:1268454033221308160}
\end{figure}

\begin{figure}[ht!]
\centering
\includegraphics[width=0.48\textwidth]{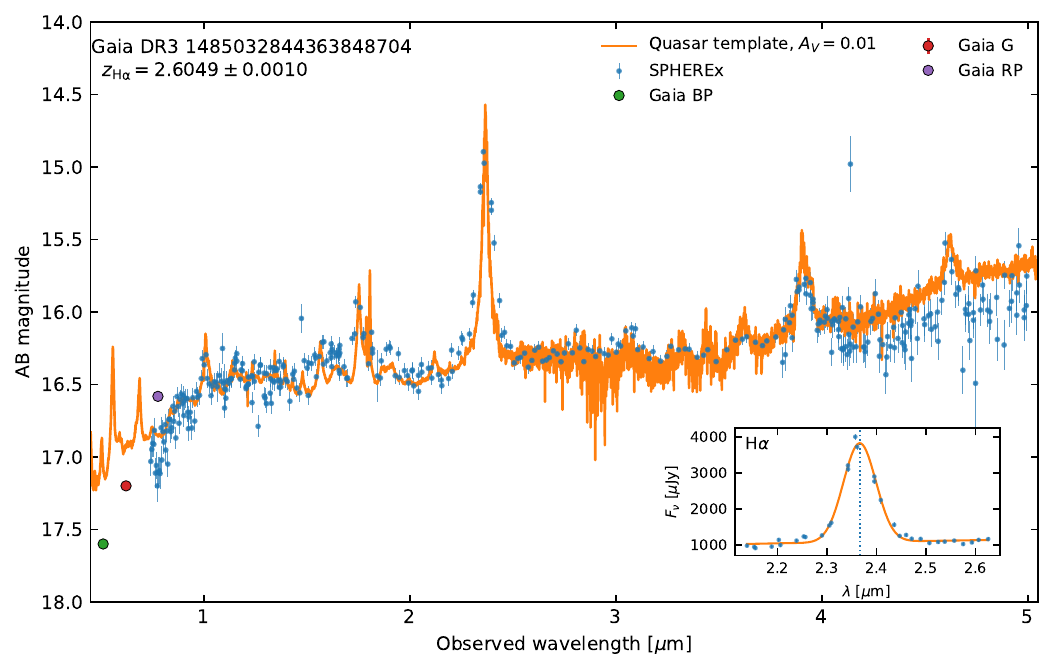}
\caption{Gaia DR3 1485032844363848704}
\label{fig:1485032844363848704}
\end{figure}

\begin{figure}[ht!]
\centering
\includegraphics[width=0.48\textwidth]{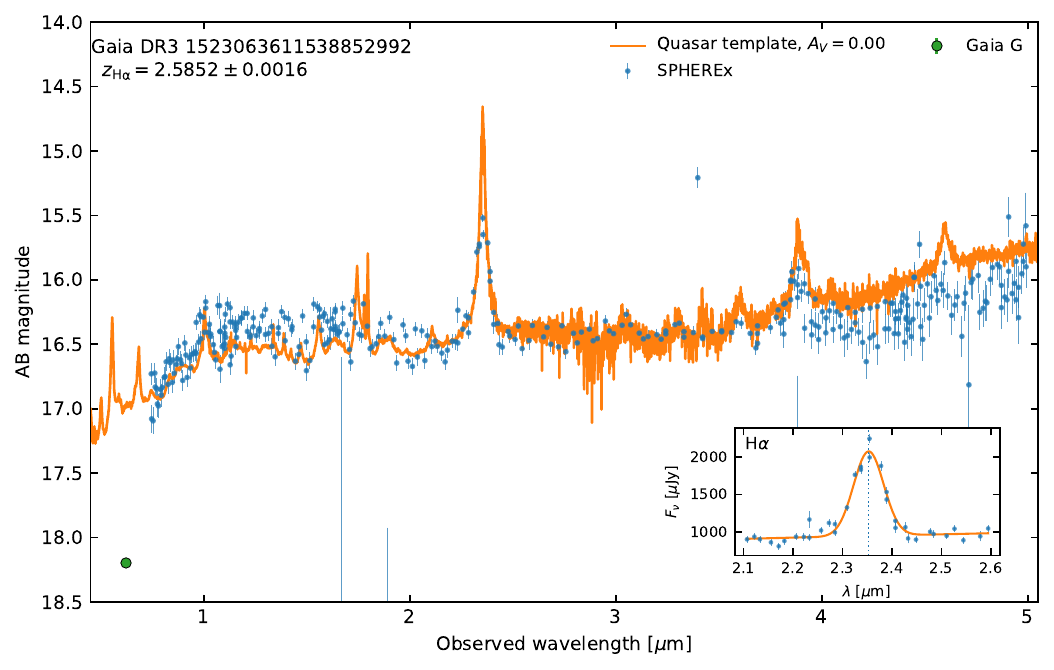}
\caption{Gaia DR3 1523063611538852992}
\label{fig:1523063611538852992}
\end{figure}

\begin{figure}[ht!]
\centering
\includegraphics[width=0.48\textwidth]{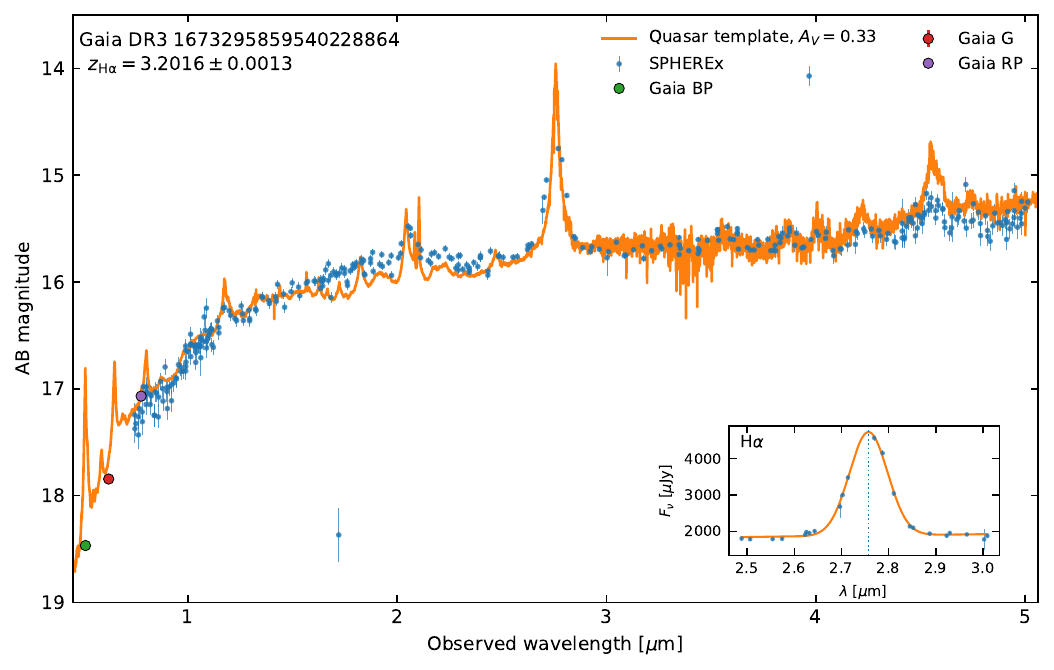}
\caption{Gaia DR3 1673295859540228864}
\label{fig:1673295859540228864}
\end{figure}

\begin{figure}[ht!]
\centering
\includegraphics[width=0.48\textwidth]{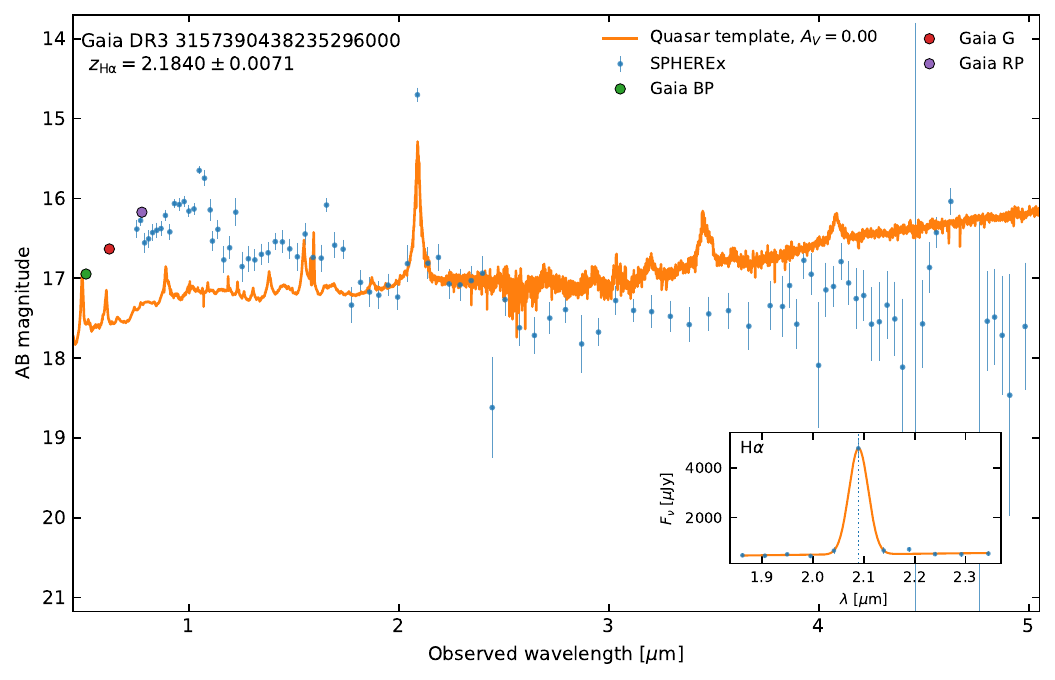}
\caption{Gaia DR3 3157390438235296000}
\label{fig:3157390438235296000}
\end{figure}

\begin{figure}[ht!]
\centering
\includegraphics[width=0.48\textwidth]{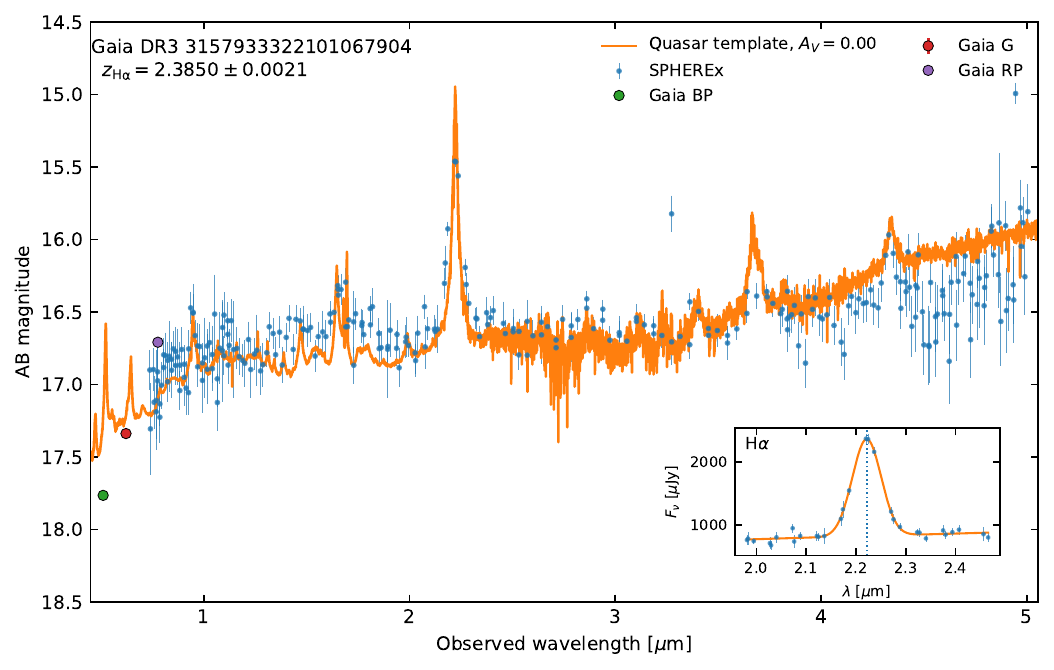}
\caption{Gaia DR3 3157933322101067904}
\label{fig:3157933322101067904}
\end{figure}

\begin{figure}[ht!]
\centering
\includegraphics[width=0.48\textwidth]{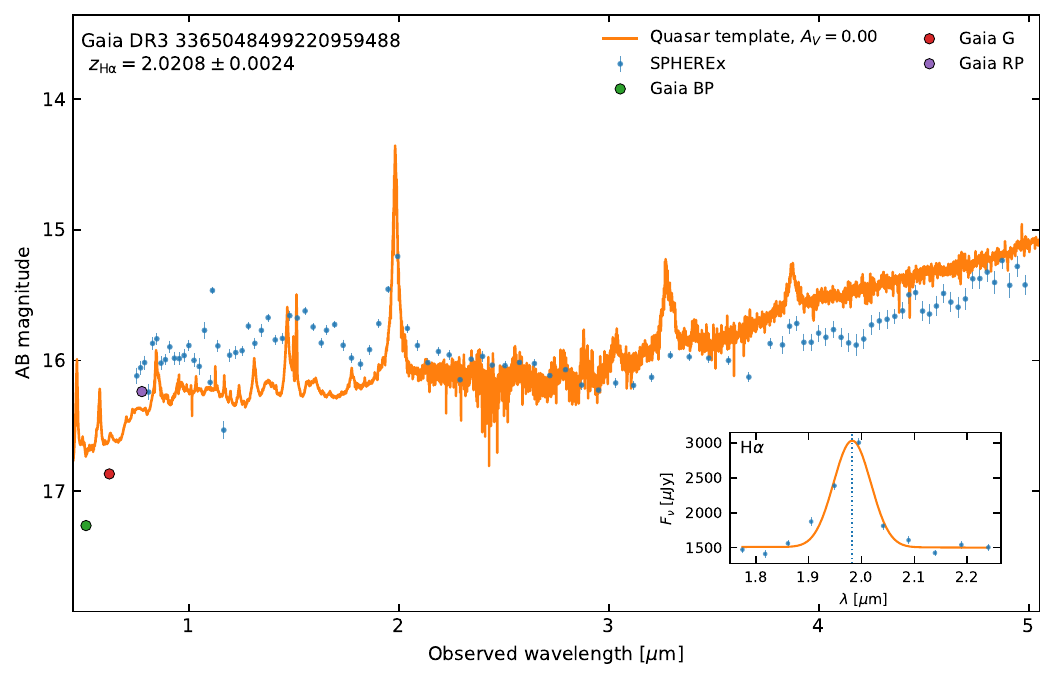}
\caption{Gaia DR3 3365048499220959488}
\label{fig:3365048499220959488}
\end{figure}

\begin{figure}[ht!]
\centering
\includegraphics[width=0.48\textwidth]{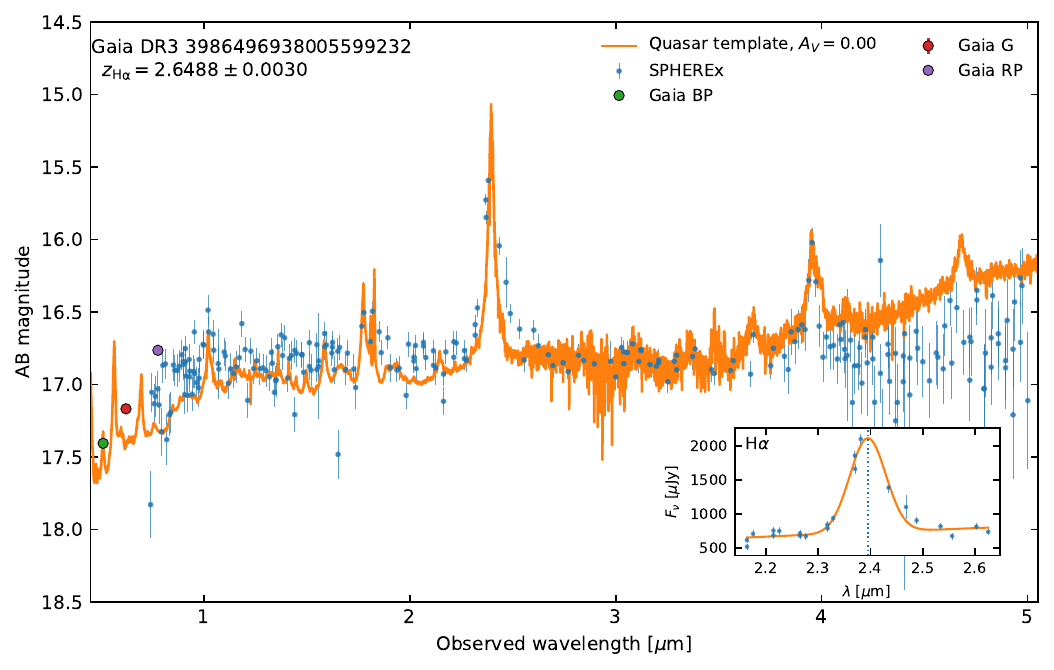}
\caption{Gaia DR3 3986496938005599232}
\label{fig:3986496938005599232}
\end{figure}
\end{appendix}

\end{document}